\RequirePackage{fix-cm}
\documentclass[natbib,smallextended]{svjour3}       
\smartqed  
\usepackage{graphicx}
\newcommand{\ssr}{{Space Science Reviews}}

 \journalname{\ssr}

\newcommand{\RJ}{\ensuremath{R_{\rm Jup}}}
\newcommand{\ME}{\ensuremath{M_{\oplus}}}
\newcommand{\RE}{\ensuremath{R_{\oplus}}}

\usepackage[dvipsnames]{xcolor}

\begin{document}

\title{Observational constraints on the formation and evolution of Neptune-class exoplanets}

\titlerunning{Neptune-class exoplanets}       

\author{Magali Deleuil  \and
        Don Pollacco \and Cl\'ement Baruteau \and Heike Rauer \and Michel Blanc
}

\authorrunning{M. Deleuil, D. Pollacco, C. Baruteau, H. Rauer \& M. Blanc} 

\institute{M. Deleuil \at
              Aix-Marseille Universit\'e, CNRS, CNES, Laboratoire d'Astrophysique de Marseille, Technop\^ole de Marseille-Etoile, 38, rue Fr\'ed\'eric Joliot-Curie
F-13388 Marseille cedex 13, France\\
              Tel.: +334-91-055929\\
              Fax: +334-91-661855\\
              \email{magali.deleuil@lam.fr}           
           \and
           D. Pollacco, \at Department of Physics
University of Warwick, Gibbet Hill Road, Coventry, CV4 7AL, 
UK 
\and 
Cl\'ement Baruteau \at
IRAP, Universit\'e de Toulouse, CNRS, UPS, F-31400 Toulouse, France
\and 
Heike Rauer \at
Institute of Planetary Research, German Aerospace Center, Berlin, Germany
\and Michel Blanc \at
IRAP, Universit\'e de Toulouse, CNRS, UPS, F-31400 Toulouse, France
}

\date{Pub Date: August 2020}

\maketitle

\begin{abstract}
Among exoplanets, the small-size population constitutes the dominant one, with a diversity of properties and compositions ranging from rocky to gas dominated envelope. While a large fraction of them have masses and radii similar to or smaller than Neptune, yet none share common properties in term of orbital period and insulation with our ice giants. These exoplanets belong to multi-planet systems where planets are closely packed within the first tenth of AU and often exposed to strong irradiation from their host star. Their formation process, subsequent evolution, and fate are still debated and trigger new developments of planet formation models. This paper reviews the characteristics and properties of this extended sample of planets with radii between $\sim$ 1.6 and 4.0 \RE. Even though we still lack real Neptune/Uranus analogues, these exoplanets provide us with key observational constraints that allow the formation of our ice giants to be placed in a more general framework than the sole example of our solar system.

\end{abstract}

\section{Introduction}
\label{intro}
The absence of detailed observational constraints, in terms of composition and atmospheric processes, on our ice giants does not allow us to fully understand their role in the evolution and final orbital architecture of the Solar System. On the other hand, exoplanets, which are now routinely detected, provide us with an enlarged sample of planets to better understand the planetary formation process.

In this paper, we review our current knowledge of exoplanets that could be seen as Neptune-class planets, their similarities and dissemblance with our ice giants, and the observational constraints they bring to current formation models. We emphasize that this class of planets does not correspond to a properly defined planet family, but because we aim at placing our ice giants in a more general context, we focus this review on exoplanets whose radius or mass are in the same range of values as those of our giants. 

Among the few thousands of exoplanets discovered to date, nearly three quarters have indeed broad characteristics comparable to those of Uranus and Neptune, i.e. radii in the range $\sim$ 1.6 to 4.0 \RE\ or masses between 10 and 20 \ME.  However, similarities are in fact still limited to their size and their mass as none of them are as far from their host star nor receives similar insolation (Sect.~\ref{overview}). In addition, the large majority belongs to packed multi-planet systems where gravitational interactions are expected to have directly affected their orbital configuration (Sect.~\ref{systems}). The diversity of these small-size planets populations, so different from those of our own solar system, challenges models of planet formation and evolution (Sect.~\ref{radii}). The location where they formed in the protoplanetary disk is still a matter of debate (Sect.~\ref{formation}). Some models even predict their building blocks might be ice-rich embryos that formed from past the snow line of their host stars \citep{Morbidelli2012,Raymond2014,Izidoro2017} but their composition remains observationally difficult to determine (Sect.~\ref{atmos}). This strengthens the need for an in-depth exploration of Uranus and Neptune, to be used as references for the exoplanet populations. While our knowledge of these systems may be incomplete, they represent a good laboratory to examine the problems of planetary formation in a broader context (Sect.~\ref{link}).

\section{Exoplanets: general overview}
\label{overview}
\begin{figure}[htbp]
\begin{center}
\includegraphics[width=0.75\textwidth]{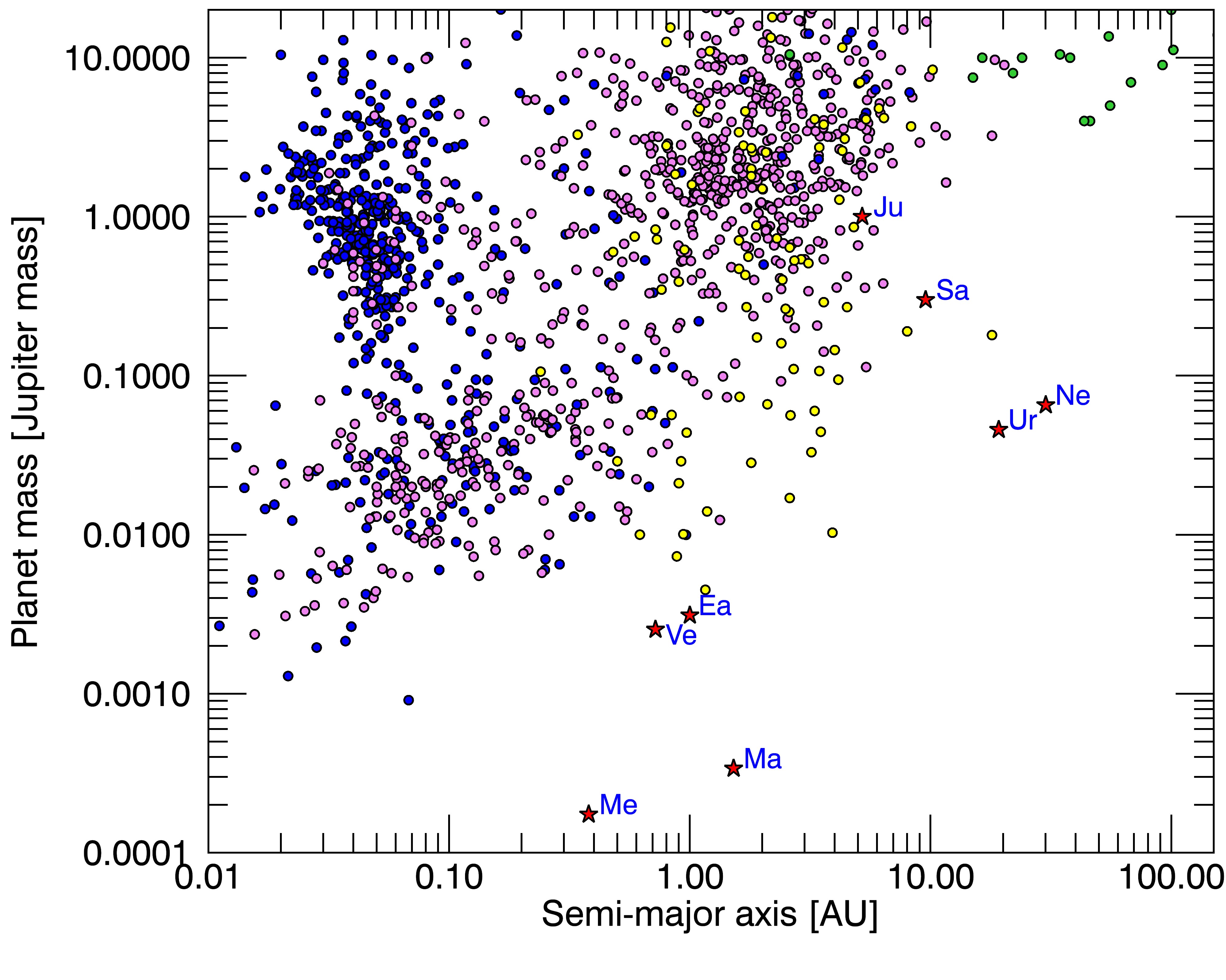}
\caption{Exoplanets with known or estimated mass as of January 2020, shown by colored circles according to their detection method: transit (blue), radial velocity (pink), microlensing (yellow) and direct imaging (green). The solar system planets are denoted by a star and the two first letters of their name. Data were downloaded from exoplanet.eu.}
\label{MassSemi}
\end{center}
\end{figure}

After more than two decades of intensive hunting, the exoplanets census amounts to $ >$4\,000 objects, with still a few additional thousands of planet candidates awaiting confirmation. These planets, which have been detected through a variety of techniques, span a large range of mass and semi-major axis parameter space (Fig.~\ref{MassSemi}). This effort has revealed planet populations with a diversity in their properties that was not expected. This surprising demography with packed planetary systems of small size planets or very massive close-in giants challenges planetary formation models, questions how our own Solar System fits into a much broader context and how typical it is. 

Despite our massive progress, exploring the exoplanet population beyond the ice line remains challenging. If more than two thirds of the known exoplanets are transiting, for orbital periods typically longer than $\simeq$ 400 days, transit surveys lack the required temporal coverage.  For such long orbital periods, the planets identification relies on the detection and analysis of mono-transits. Moreover, the geometric transit probability of these long period objects is extremely low, making the observation of their transit a very low probability event. The sole validated transiting Jupiter analog so far remains Kepler-167e  \citep{Kipping2016}.  With a radius of 0.91 $\pm$ 0.02 \RJ, it orbits its K4-dwarf host with an orbital period of 1071.2323 $\pm$ 0.0006 d and receives a stellar isolation close to Jupiter's. While this Jupiter analog is well beyond the snowline of its host star, we lack an estimate of its mass which is needed to understand its composition.

Radial velocity surveys have gathered observations over some two decades. While in principal this may seem sufficient to identify the Doppler signatures of ice giants, the velocity amplitudes of these relatively long period planets often require more than a hundred measurements over a 10-15 year time span to cover a complete planetary orbital phase or at least to safely identify a long-term trend in radial velocity measurements.

Nonetheless, the population of gas giant planets with masses, orbits, and eccentricities similar to those of Jupiter's now amounts to  a few tens of planets \citep{Boisse2012,Wittenmyer2014,Rowan2016}, allowing rough estimates of the population properties and occurrence rates. Current estimates of the frequency of Jupiter analogs range from 6.2$^{+2.8}_{-1.6}$\% \citep{Wittenmyer2016}  down to  $\simeq$ 3\% \citep{Rowan2016} - a range that may reflect inadequacies in our definition of analogs  amongst other possibilities.

Estimates of the planetary occurrence rates remain indeed approximate due to uncertainties on the underlying stellar populations, unknowns such as the intrinsic planetary systems architecture, biases in the surveys completeness, and various sources of systematic errors that directly affect the results. Other issues that affect these estimates include the possible influence of outer planets on the formation and evolution of interior planets. This has been investigated either through theoretical modeling \citep{Childs2019} or observational uncertainty e.g. the search for long term trends in systems hosting at least one super-Earth planet \citep{Bryan2019,Zhu2018}. Although there is some disagreement over the derived estimates which range from 39 $\pm$ 7\% to $\sim$ 90\%, they consistently suggest a high occurrence rate of long-period gas giants in systems with super-Earths.  

Beyond Saturn, the mass and semi-major axis parameter space of ice analogs largely exceeds the radial velocity surveys sensitivity and others techniques are more appropriate. Microlensing surveys have reported a few cold planets with masses comparable to that of the solar system ice giants \citep{Poleski2014, Poleski2018}. While they have the required sensitivity to probe this mass - period domain, the non-repetitiveness of the event, the degeneracies of both the lens and the source, but mostly the faintness of the host star, make the techniques more suited for statistical inferences of cold planets occurrence \citep{Penny2019} but not any further detailed characterization. 

For nearby stars, the direct imaging and astrometric techniques are well suited to produce data on long period planets. In the case of imaging we are limited to the most luminous, i.e. youngest, planets. Astrometric results can also be used alongside  radial velocity data and can be most powerful for those bright radial velocity targets that have been monitored for decades. Combining these methods therefore has the potential to explore the domain of intermediate semi-major axes \citep{Boehle2019,Kane2019}, especially in the GAIA era.

\section{Orbital period ratios in compact multi-planetary systems}
\label{systems}
\begin{figure}[t]
\begin{center}
\includegraphics[width=0.75\textwidth]{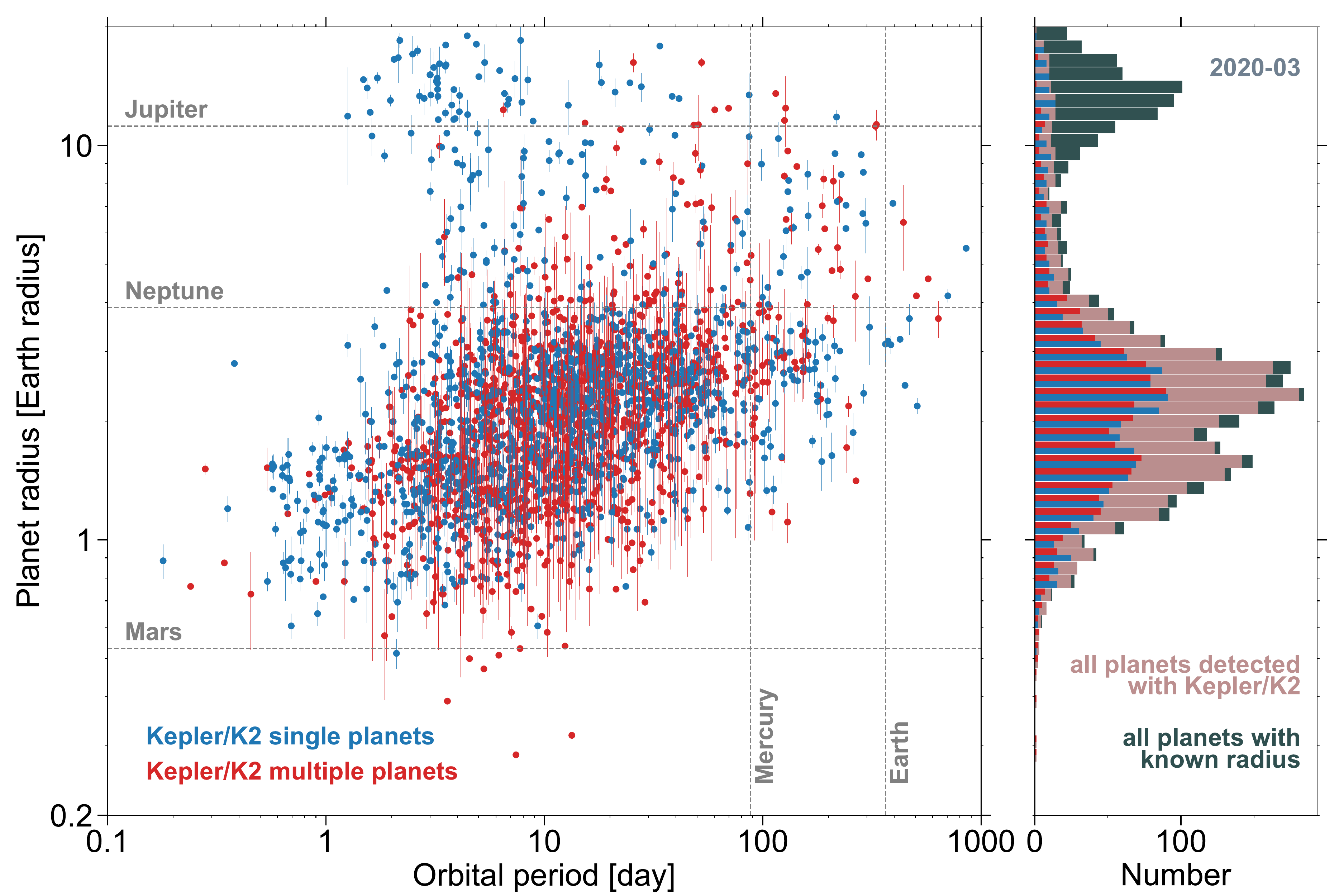}
\caption{Physical radius and orbital period of the planets detected by transit with the Kepler space telescope (2404 planets whose status is confirmed, including those of the K2 campaign). Planets that are part of multi-planetary systems are shown in red, single planets are in blue. 1$\sigma$ error bars are overplotted. The right part of the figure compares the distribution of physical radii of the Kepler planets with that of all planets with known radius (gray histogram, 3083 planets). Data were downloaded from exoplanet.eu. }
\label{RadDistrib}
\end{center}
\end{figure}

Over the last decade, the Kepler/K2 missions have dramatically increased the number of small planets known, i.e. with physical radius $\leq 4$\RE, and these planets now dominate the planet population overall (Fig.~\ref{RadDistrib}). The large number and diversity of planets similar to and smaller than Neptune was not a robust prediction of planet formation theories before these missions, partly due to the uncertain rate of orbital migration for this class of planets at the time \citep[see, e.g., the review by][]{Benz14}.

\begin{figure}[htbp]
\begin{center}
\includegraphics[width=0.65\textwidth]{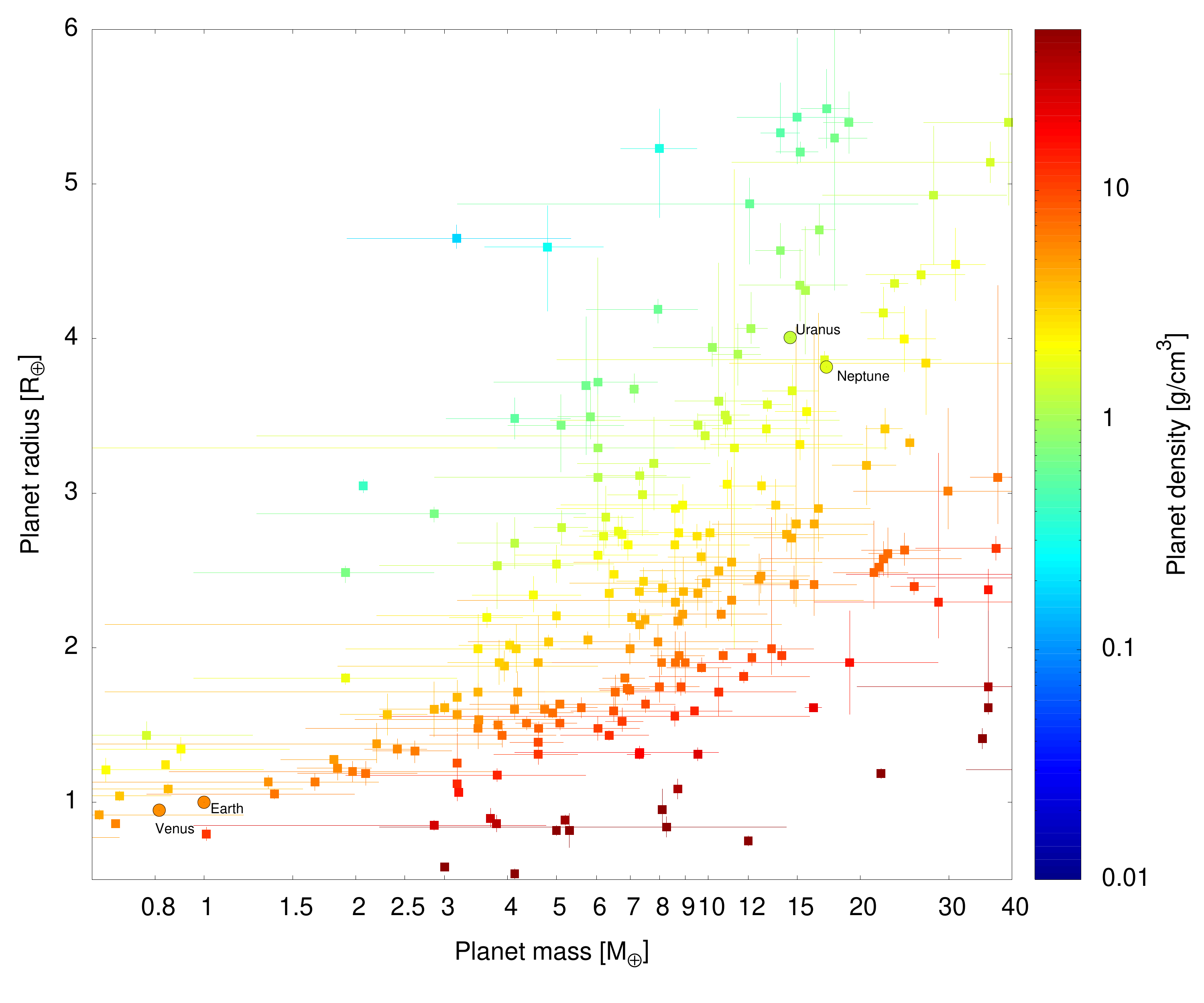}
\includegraphics[width=0.66\textwidth]{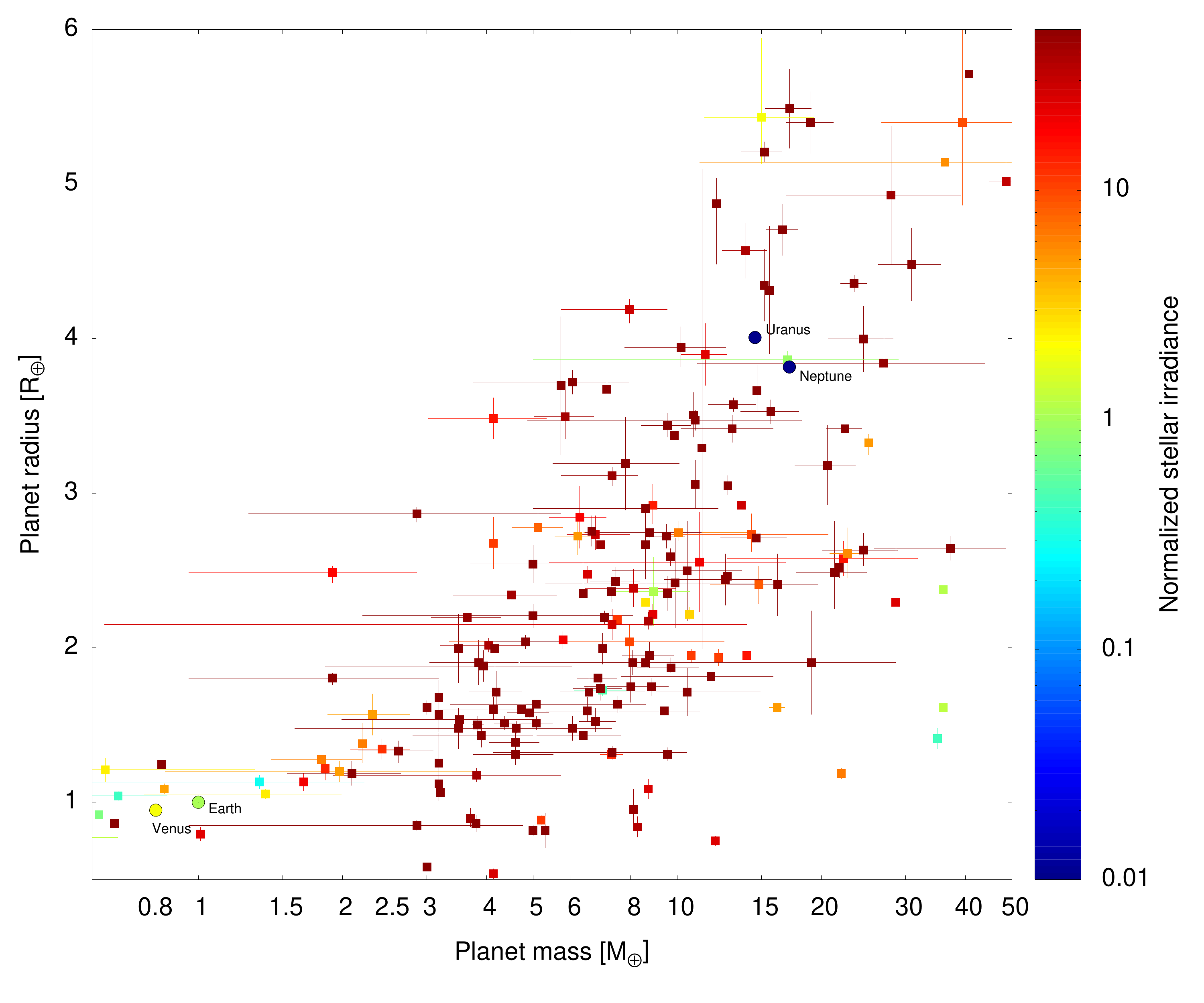}
\caption{Mass - radius diagram for exoplanets with masses in the 0.8-50 \ME\ range. The marks' color scales with planets density (up) and with stellar irradiation normalized to the Solar one (bottom) following the colorbar on the right-hand side.
Data were downloaded from exoplanet.eu.}
\label{MassRadius}
\end{center}
\end{figure}

Their physical and dynamical properties are indeed at odds with planets of similar size in our Solar system. This is illustrated for instance in Fig.~\ref{MassRadius}, which shows the radius versus the mass of the exoplanets in the range between 0.8 and 50~\ME\, as well as the density (upper panel) and stellar irradiation (lower panel), and how they compare to Venus, Earth, Uranus, and Neptune.

About half of these transiting planets belong to flat and compact multi-planet systems, with up to 8 planets \citep{Lissauer2011, Fabrycky2014,  Santerne2019}. Note that in such packed systems, gravitational perturbations are important and may induce, for consecutive transits, a significant and measurable deviation from strictly periodic transit times. These transit time variations (TTV) can provide another way to estimate planetary masses through dynamical modeling \citep{Barros2015,Jontof-Hutter2015}. However, uncertainties on the real number of planets in the system, the good phase coverage and long baseline needed, and the complexity of the modeling prevent precise estimates of planetary mass \citep[e.g.,][]{Weiss2013} so that, in the best case, they remain affected by large uncertainties.

\begin{figure}[htbp]
\begin{center}
\includegraphics[width=0.75\textwidth]{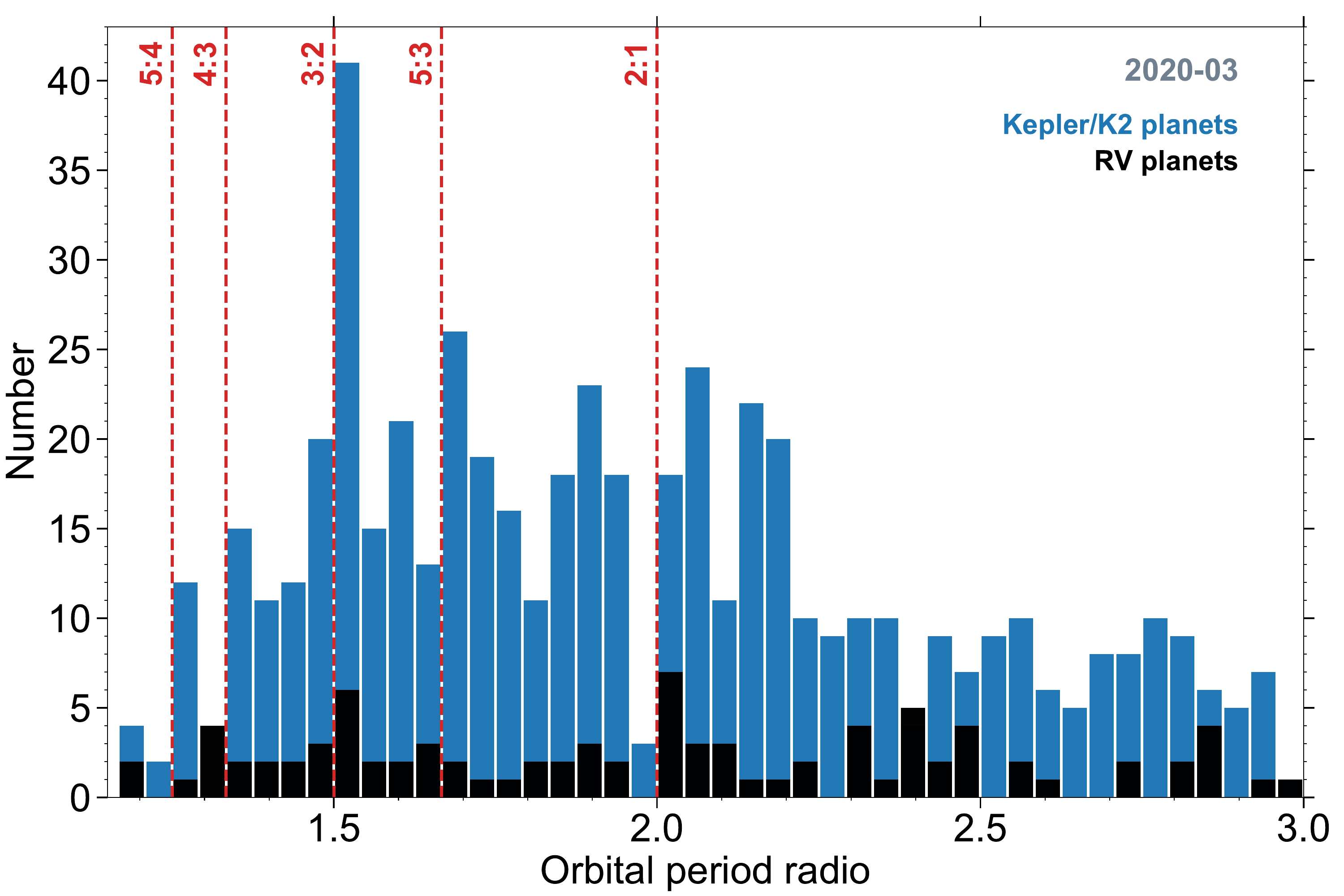}
\caption{Distribution of the orbital period ratio for pairs of adjacent planets among the multi-planetary systems detected by Kepler and K2 (in blue) and by radial velocity (in black). Bins are 1/24 wide. Vertical dashed lines show the location of a few mean-motion resonances. Data were dowloaded from exoplanet.eu.}
\label{opr}
\end{center}
\end{figure}

The architecture of multi-planetary systems may help constrain planet formation and evolution scenarios. In this regard, an interesting quantity is the orbital period ratio for pairs of adjacent planets in compact multi-planet systems \citep[see, e.g.,][]{Steffen2015}. Its distribution is displayed in Fig.~\ref{opr} for the confirmed planets detected by Kepler and K2 (in blue) and for the planets detected by radial velocity (in black). Both distributions show a large and surprising diversity, with the overall trend that adjacent planet pairs near mean-motion resonances tend to have period ratios slightly greater than resonant values. This is particularly clear near the 3:2 and the 2:1 mean-motion resonances, even for radial velocity planets, despite a smaller statistics.

The diversity of orbital period ratios among the close planet pairs detected by Kepler/K2 has stimulated many theoretical works. In particular, it is often thought that two planets undergoing convergent migration in their protoplanetary disk should necessarily end up in mean-motion resonance. Turbulence in the inner parts of protoplanetary disks could actually explain why many pairs of low-mass planets are either near- or non-resonant \citep[e.g.,][]{Pierens11,Paardekooper13}. Also, interactions between planets and the wakes of their companions in the disk could explain why many pairs of planets in the super-Earth to Neptune size range have period ratios slightly greater than resonant \citep{Baruteau13}. This feature could alternatively arise from star-planet tidal interactions for the closest-in planet pairs \citep[see, e.g.,][]{Papaloizou11}. Other studies have gone a different route and examined the possibility that super-Earths and warm-Neptunes could have formed mainly by {\it in-situ} growth after the dissipation of the inner protoplanetary disk \citep{Hansen13,Ogihara2015}.

\section{Distribution of physical radii and dependence on stellar metallicity}
\label{radii}
The mean density can be used to probe the composition of planets but, to be meaningful, it requires accurate estimates of planetary radii and masses. With current instruments these observations remain a challenging task and, consequently, only a handful of small planets with well-measured masses, to a precision better than 20\% \citep[e.g.][]{Motalebi2015, Armstrong2020} are known. This arises because the host stars of the first generation of transiting surveys are generally faint and the amplitude of the expected radial velocity signal of small size planets is low ($<$ 1 m/s), that is well below the photon noise in the high magnitudes regime. Hence, we are often left with just an upper limit on the planetary mass.

Planetary radii measurements can be determined with greater accuracy than planetary masses provided the host star parameters are accurately known. For periods shorter than 100 days, the radii distribution displays a bimodal shape (see Fig.~\ref{RadDistrib}), highlighting two planets populations with typical sizes of $\simeq$ 1.3~\RE\ (super-Earths) and $\simeq$ 2.4~\RE\ (sub-Neptunes), with a scarcity of planets around 1.8~\RE\ \citep{Owen2013, Fulton2017,Fulton2018}.  This was previously suspected from the analysis of mean densities of some Kepler planets with masses estimates \citep{Weiss2014,Rogers2015}. These studies suggested a population of rocky planets whose density increases with increasing radius up to $\simeq$ 1.5\RE, and above, another population of planets whose density rapidly decreases with increasing radius, compatible with planets with large gaseous envelopes overlying a rocky core. However, as planets in this regime have inaccurate mass estimate, it is difficult to recognise any trends in their composition (or diversity;  see Fig.~\ref{MassRadius}).

Various processes, related to different aspects of planets formation or to evolution mechanisms, are invoked to explain this bimodal distribution: different compositions of solid cores, the accretion in gas-rich or gas-poor environments, or the loss of gaseous envelopes due to photo-evaporation. The photo-evaporation scenario is supported by observational evidences \citep{VanEylen2018} of a clean gap between the two populations, whose position decreases with increasing orbital period, but also by synthetic planet populations in the presence of irradiation \citep{Owen2018,Jin2018}. In this case, super-Earths could be the cores of planets with an initially relatively low mass ($<$ 20 ~\ME) but with a large ice content, which eventually become bare rocky cores due to efficient evaporation. The rocky composition suggests that these planets have accreted mainly inside the ice line and with migration confined to the inner parts of their protoplanetary disk.  Whether this migration covered several or only a fraction of AU from their current locations remains to be established.

The frequency of planets and its dependency with the host star metallicity brings some further insights into the planets formation mechanism. For radial velocity planets, \cite{Courcol2016} suggested that the frequency of exo-Neptunes (between 10 and 40~\ME) is correlated with the host star metallicity, similar to that found for giant planets. For Super-Earths ($<$ 10~\ME) this correlation does not hold. This has been confirmed by \cite{Petigura2018}, who have shown from Kepler planets that, while warm super-Earths occurrence is nearly constant over metallicities, the occurrence of warm sub-Neptunes doubles over that same metallicity interval, from 20 to 40 planets per 100 stars. This increased occurrence rate of hot Neptunes around metal-rich stars could be seen as evidence for high-eccentricity migration of the planets or metallicity dependent photo-evaporation \citep{Owen2018}.

More recently, alternative hypotheses have been explored to overcome difficulties in removing a giant planet's entire H/He atmosphere, even in presence of strong irradiation, such as a core-powered mass-loss \citep{Ginzburg2018, Gupta2019, Gupta2020}, or two distinct formation pathways for the two sub-populations \citep{Zeng2019, Mousis2020}. Subject to high-energy radiation from their host star, close-in exoplanets are expected to experience atmospheric escape \citep{Lammer2003}. Observational evidence of gas escaping at high velocities has been reported for two hot Neptunes, {\bf GJ 436 b} and GJ 3470 b \citep{Ehrenreich2015,Bourrier2018} observed through the detection of hydrogen absorption signatures in the Ly-$\alpha$ line of their host star. As a result, the planets appear surrounded by an extended upper atmosphere of neutral hydrogen. In the extreme case of GJ 436 b it absorbs more than half of its M-type star's Ly-$\alpha$ emission \citep{Bourrier2015} and extends over half of the planet's orbit \citep{Lavie2017, dosSantos2019}. The lack of a well-defined stellar continuum and the strong interstellar medium absorption that affects the Ly-$\alpha$ emission profile prevent a precise determination of the inferred mass loss. Current estimates show that photo-evaporation might be of a few \% only \citep{dosSantos2020} but could be up to 35\% in the case of GJ 3470 b, depending on the XUV and Lyman $\alpha$ stellar irradiation but also the past evolution of the planet.
A possible negligible effect of photo-evaporation on the planets composition is confirmed at least in one case, by the detection from the ground and from space of the metastable helium triplet at 10833 \AA\  in HAT-P-11b \citep{Allart2018, Mansfield2018}. Although HAT-P-11b is among the most massive of its class, these observations suggest the bulk composition of close-in exo-Neptunes could remain mostly unaffected during their evolution.

\section{Atmospheric composition}
\label{atmos}
With their large atmospheric scale height, Neptune-class planets appear as the best targets for atmospheric characterization among the small-sized planet population. Their atmosphere properties can shed light on their composition, and give us hints on how and where in the protoplanetary disk they formed as well as their subsequent evolution. In particular, the metallicity is a key factor that could help disentangle between different formation paths. If close-in exo-Neptunes were formed beyond the ice line by accretion of water-rich planetesimals and experienced subsequent migration, they should have much higher metallicities and their atmosphere should be H$_2$O enriched. In contrast to low temperature ice giants in which oxygen is trapped deep in the envelope \citep{Atreya2020}, in exo-Neptunes water can be used as an indicator of the oxygen abundance and, hence, of the metallicity. 

\begin{figure}[htbp]
\begin{center}
\includegraphics[width=0.55\textwidth]{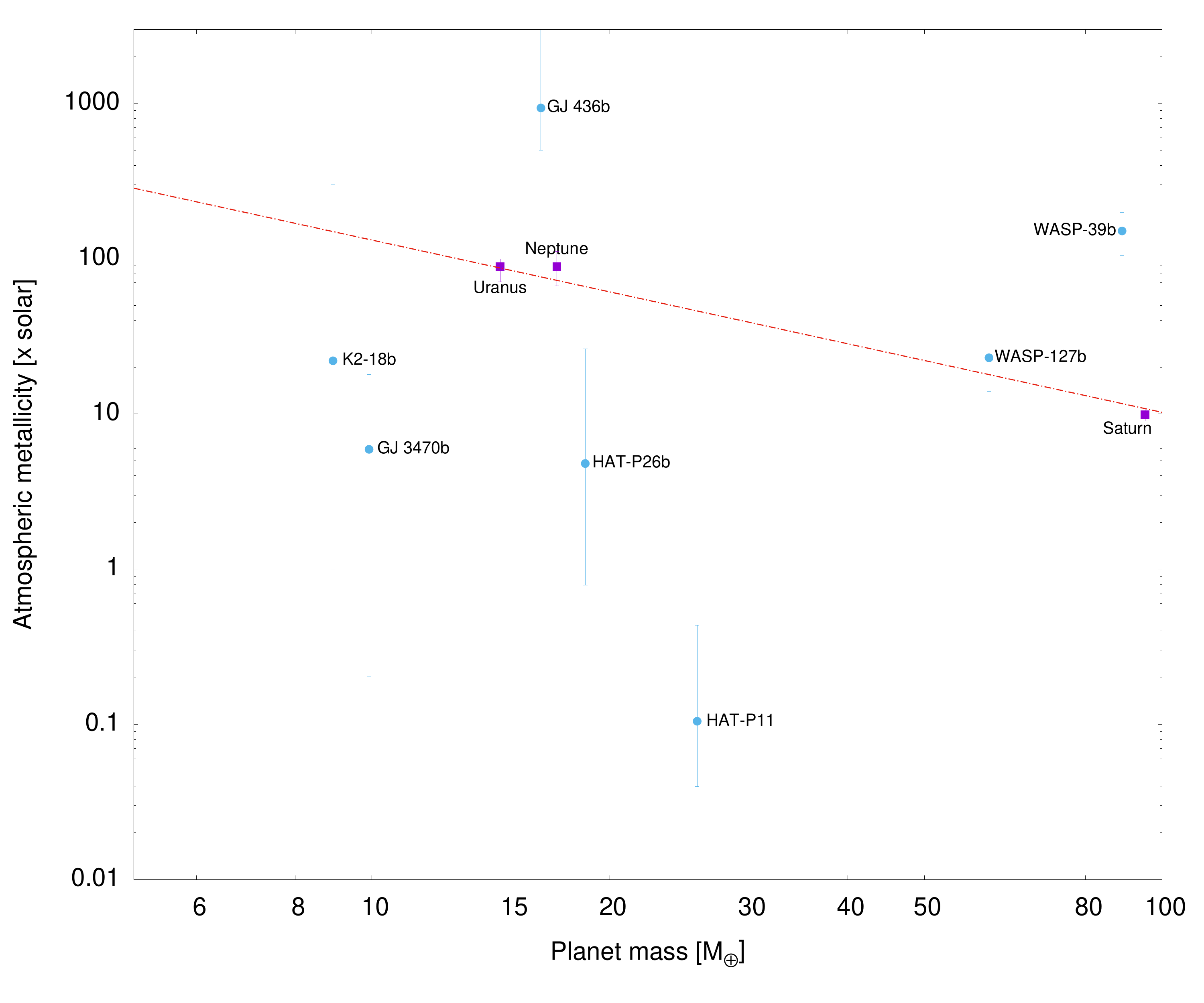}
\caption{Mass-metallicity diagram in the 5 - 100 \ME domain. The metallicity proxy for solar system planets is CH$_4$  but H$_2$O for the exoplanets. The dashed line is the linear fit derived for the solar system giants (adapted from \citealp{Wakeford2018}).}
\label{metallicity}
\end{center}
\end{figure}

In recent years several atomic and molecular species have been detected in the transmission spectra of exo-Neptunes (see Table~\ref{species} and references therein). Most of these detections have been performed in the near infrared with Hubble. The WFC3 wavelength ranges are not the best suited for the main carbon bearing species, and water vapor, via its absorption feature at 1.4$\mu$m, is the molecule that is measured in the atmosphere of all these planets. In itself, H$_2$O is a key molecule as its abundances offer insights into the formation history, bringing indirect constrains on the physical properties of the original planet's building blocks. In principle, it could indeed allow to differentiate between accretion of water-rich planetesimals beyond the ice line or in-situ or at least formation inward of the ice line. In these warm-to-hot atmospheres, the amplitude of the H$_2$O feature is attenuated but current measurements rule out clear solar metallicity atmospheres. 
In addition to measurements or estimates of the abundances of chemical species, atmospheric retrieval techniques have enabled to estimate their metallicity (see Fig.~\ref{metallicity}). While the atmospheric C/H ratio, derived from CH$_4$ abundances, is used as a proxy for the metallicity of the solar system giants, in exoplanets, it is usually based on retrieved water abundances alone and some assumptions on the basic ingredients available, the atmospheric chemistry, and the clouds distribution \citep[see, e.g.,][]{Madhusudhan2016}. 
 Current chemical compositions point to low metallicities in their atmosphere in contrast to the super-solar metallicities of Neptune and Uranus, the latter seen as resulting from accretion of H$_2$O-rich planetesimals (Fig.~\ref{metallicity}) and \cite[e.g.][]{Welbanks2019}.
 \begin{table}[ht]
\caption{Exoplanets in the 5 - 25 \ME\, range with detected atmospheric species in the published literature.}
\begin{center}
\begin{tabular}{|p{1.8cm}||p{1.8cm}|p{1.8cm}|p{3.2cm}|p{1.5cm}|  }
\hline
Planet Name & Radius & Mass & Measured species & References \\
                      &   [\RE]    &   [\ME]  &                               &                   \\
\hline
\hline
K2-18 b & 2.61 $\pm$ 0.09 & 8.63 $\pm$ 1.35 & CH$_4$, H$_2$O, NH$_3$ & 1,2,3 \\
GJ 3470 b & 4.19$\pm$ 0.59 & 12.58$^{+1.31}_{-1.28}$ & CO, H, H$_2$O, He & 4,5,6,7 \\  
HAT-P-26 b & 6.33 $\pm$ 0.58  & 18.6 $\pm$ 2.3  & CH$_4$, CO, CO$_2$, CrH, H$_2$O, ScH, TiH & 8,9\\
HAT-P-11 b & 4.36 $\pm$ 0.06  & 23.4 $\pm$ 1.5  & He, H$_2$O &  10,11  \\ 	
GJ 436 b & 4.8 $\pm$ 0.2 & 13.9$\pm$ 1.5 &  CH$_4$, CO, CO$_2$, H, H$_2$O, Si & 12  \\
\hline
\end{tabular}
\end{center}
{References: (1) \cite{Madhu2020} , (2) \cite{Benneke2019}, (3) \citep{Tsiaras2019}, (4) \cite{Benneke2019}, (5) \cite{Bourrier2018}, 6 \citep{Palle2020}, (7) \cite{Ninan2019}, (8) \cite{MacDo2019}, (9) \cite{Wakeford2017}, (10) \cite{Welbanks2019}, (11) \cite{Allart2018}, 
(12) \cite{Morley2017},  
}
\label{species}
\end{table}%
This could be seen as a large diversity in composition resulting from different formation pathways or differences in disks composition. However, there is a degeneracy between clouds and gas abundances, especially when analyses are done over a very limited spectral coverage \citep[see][e.g.]{Benneke2019}. In addition, a proper comparison between exo-Neptunes and our solar system giants would deserve not only to enlarge the sample of planets with atmosphere studies, but mainly a proper inventory of key species in their atmosphere. This also further enhances the need to improve our understanding of the atmospheric composition of Neptune and Uranus.

As a matter of fact, these measurements are related to species in the very upper parts of the atmosphere of these unresolved objects. In addition, transmission spectroscopy probes different atmospheric layers along the line of sight, layers which are at different heights but also on the day or the night sides. This implies a strong heterogeneity along the line of sight which is difficult to account for in the atmosphere composition retrieval  \citep{Caldas2018}. In addition, these measurements come from different models which use different assumptions,  including the possible presence of hazes and clouds, and different compositions resulting on an additional difficulty when trying to compare the inferred results \citep{Benneke2019, Wong2020}.  

Finally, compared to our ice giants, these planets are highly irradiated. As a consequence, we expect differences in their thermal structure, the main source of heating being located outside the atmosphere rather than in the interior of the planet, with a vigorous day-night heat distribution \citep{Spiegel2010, Parmentier2018}.  In addition, the interior of these planets remains mostly unknown: whether we are facing low density fluid objects or a solid differentiated body topped by a thick atmosphere remains to be well established \citep[see, e.g.,][]{Madhu2020}. For the special case of fluid objects, the transport and mixing of material between different layers \citep{Parmentier2013, Zhang2018, Komacek2019} can also affect the atmospheric chemistry and the cloud formation, making the atmospheric composition not well representative of the planet's composition.

\section{Formation and orbital evolution}
\label{formation}
The formation and orbital evolution of planets up to Neptune's size has been extensively examined within the core-accretion model with core growth via planetesimal accretion and orbital migration via interactions with the protoplanetary disk. The growth timescale by planetesimal accretion beyond the  ice line can be longer than the typical lifetime of protoplanetary disks and the migration timescale of sub-Neptune planetary cores, which makes it difficult to account for the giants in the Solar System \citep[see, e.g.,][and references therein]{Bitsch15}.

This difficulty has allowed a new paradigm to emerge for core accretion, where growth is dominated by the accretion of sub-meter-sized solids in the protoplanetary disk that are commonly referred to as pebbles. Pebbles interact much more strongly with the disk gas than planetesimals indeed, which allows the gas to damp their eccentricity and inclination much more efficiently as they zip through the vicinity of a planetary core \citep{Ormel10}. Although much less massive than planetesimals, pebbles are therefore accreted much more efficiently, which can largely speed up planetary growth, particularly beyond the ice line \citep{Lambrechts14}. In particular, the formation and evolution model of \citet{Bitsch15}, which includes pebble accretion and planet migration in a time-evolving, one-dimensional protoplanetary disk, shows that it may be possible to form ice giants (greater than 2$M_{\oplus}$) both in the inner and outer parts of the disk, if such planets form quite late in the disk evolution (after a few million years). Global models of this kind still have room for improvements. This includes taking into account gas accretion onto growing planetary cores, to better understand the conditions for Neptune-mass planets to acquire a gaseous envelope with a similar mass ratio as the ice giants in the Solar System \citep[see, e.g.,][and references therein]{Venturini17,Lambrechts19}. Collisions between planetary cores during the dissipation of the protoplanetary disk could also play a prominent role in the emergence of Neptune-like planets, and \citet{Izidoro15} have shown that the final growth and evolution of our ice giants could have proceeded by collisions between several planetary embryos of a few Earth masses beyond Saturn's orbit. Such a process could also account for their high obliquity.

We finally mention an alternative, plausible pathway to forming Neptune-mass planets via the so-called tidal downsizing scenario.  In this scenario, gravitational instabilities occuring beyond 50 to100 au from the star, would cause the outer parts of the protoplanetary disk to fragment. The disk fragmentation forms gas clumps with a mass comparable to that of Jupiter, which migrate inwards rapidly due to their interaction with the disk gas. The inspiraling clumps may build a solid core, and may experience substantial tidal disruption by tides from the central star. The tidal downsizing scenario may form (sub-)Neptune-mass planets with very diverse internal and orbital properties, and we refer the reader to \citet{NayakDawes} for a thorough review on this formation scenario.
\\

\section{Towards connecting exo-Neptunes and ice giants}
\label{link}

\begin{figure}[ht]
\begin{center}
\includegraphics[width=1.0\textwidth]{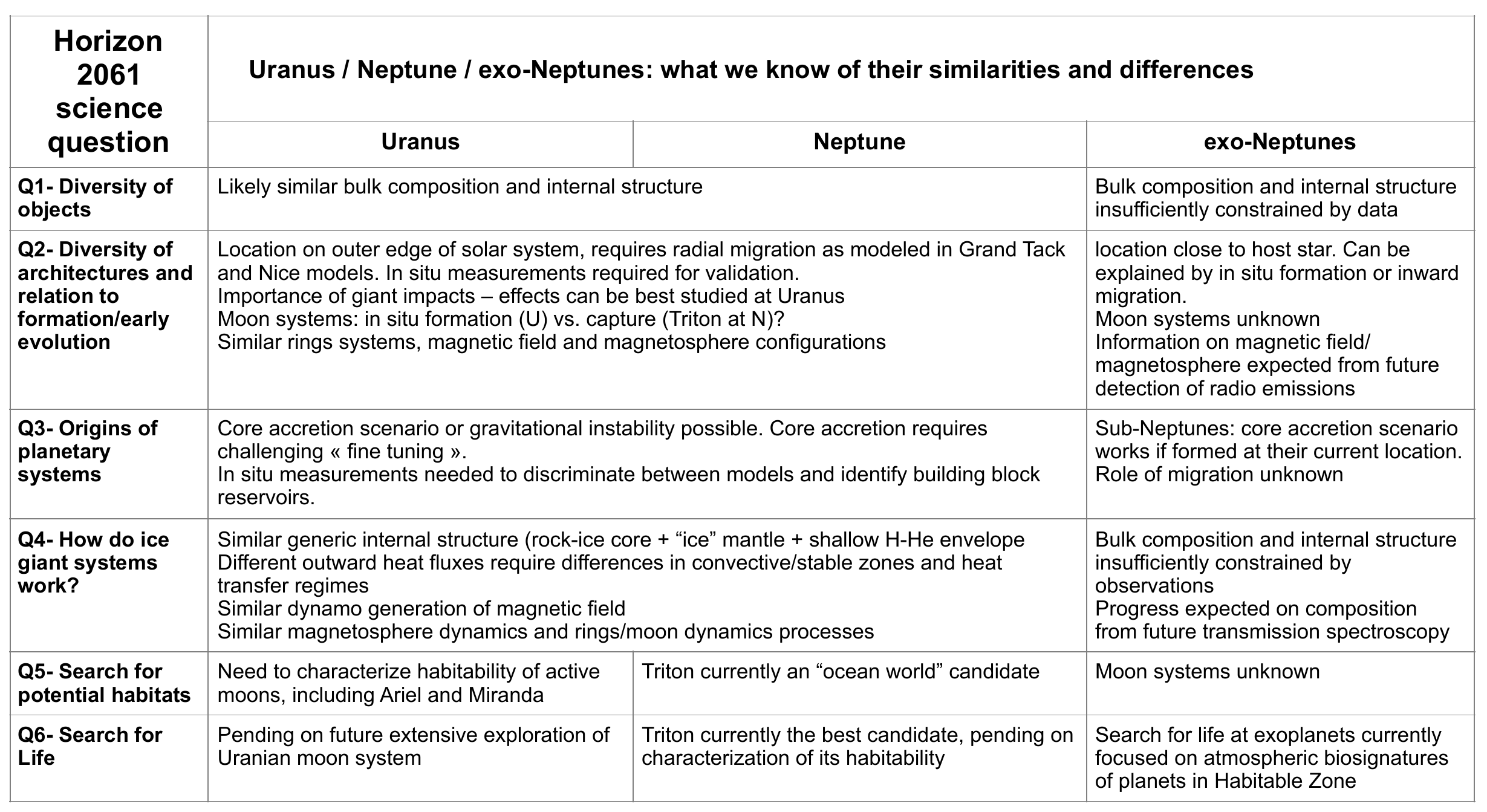}
\caption{A tentative survey of known similarities and differences between Uranus, Neptune and exo-Neptunes, established on the basis of our current knowledge and in the light of the six “key science questions” of the “Planetary Exploration, Horizon 2061” foresight exercise (https://horizon2061.cnrs.fr). From \cite{Blanc2021}, this issue.
}
\label{SimilaritiesDiff}
\end{center}
\end{figure}

We have reviewed the properties of this broad class of exoplanets with radii and masses slightly smaller than those of Uranus and Neptune, highlighted in Fig.~\ref{RadDistrib}
On the basis of the current inventory of exoplanets, this Neptune-class planets are not directly comparable under all aspects to solar system ice giants, in the first place because they are located on orbits with significantly shorter periods, e.g. much closer to their stars. In order to better assess how Uranus, Neptune, and these exo-Neptunes are similar, and how they differ, the table shown in Fig.~\ref{SimilaritiesDiff}, reproduced from \cite{Blanc2021}, offers a concise summary of their similarities and differences in light of the six key scientific questions of the Horizon 2061 foresight exercise, based on our current knowledge. 

Focusing on Question 1 (understanding the diversity of objects), the similar mass and radius ranges of exo-Neptunes and ice giants suggest similar densities, bulk compositions and therefore ice-to rock ratios. It is in this restricted sense only that one can provisionally say today that ice giants and exo-Neptunes belong to the same class of planets. But it is only by comparing their characteristics more accurately in the decades to come, as observation techniques will make progress, that we will really be able to understand whether they belong to the same unique class or if they are representative of two different classes in terms of planetary composition and evolution. To achieve this goal, a major quantitative step forward in our knowledge of Neptune-class planets will have to be accomplished. This effort needs to be made in two directions:
\begin{itemize}
\item[$\bullet$] For exoplanet research, one will need to take advantage of progress expected in the decades to come in the panoply of ground-based and space-based techniques for detecting and characterizing exoplanets, to (i) extend the survey of exoplanets towards the right-hand side of Fig.~\ref{MassSemi}, with the objective of ultimately covering the range of masses/sizes and orbital periods where Uranus and Neptune reside;  (ii) gain orders of magnitude in the characterization of the densities and atmospheric compositions of Neptune-like planets. 
\medskip
\item[$\bullet$] For solar system research, one needs to develop the necessary tools to provide orders-of-magnitude improvements in our poor knowledge of ice giants. Such progress can be accomplished only by sending space missions to explore ice giants with a combination of atmospheric entry probes and low-periapse high-inclination orbiters (Blanc et al., 2020). Atmospheric entry probes of such urgently needed ice giants exploration program, will provide accurate measurements of noble gases chemical composition and key isotopic ratios of ice giants atmospheres \citep{Mousis2018, Mandt2020} while orbiter measurements will provide complementary measurements of heavy elements in the atmosphere and of the internal structure, “à la Juno”  \citep{Bolton2017}, via a spectacular improvement of the determination of gravity and magnetic fields. This combination of probe and orbiter measurements will make it possible to determine at the same time the bulk and atmospheric compositions of ice giants, making a direct and accurate comparison with Neptunes-class exoplanets possible for the first time. 
\end{itemize}

\section{Conclusion}
There is no clear Neptune analogs identified today and likely we will have to wait for the next generation of imagers to catch them.  Despite developments in imaging are mostly focused on the detection of Earth analogs, detecting real ice giant analogs would be a major step in our understanding of the planetary system architecture.

Meanwhile, we are left with planetary systems very different from ours. Here, Neptune-class planets are found in closely packed multi-planet systems and all are close to their host star (<0.1AU). Exposed to strong irradiation from their host star, suspected of being able to strip them from their atmosphere, some might have ended up as naked cores. If this scenario is confirmed, exo-Neptunes would be the progenitors of the super-Earth population, and therefore the most common product of planet formation.

In the JWST era, these transiting exo-Neptunes with their extended atmosphere are key targets for detailed characterization of their atmospheric composition, which will likely bring new insights and constraints of their formation mechanism. First hints into their atmospheric properties suggest however a large diversity at odds to that of the ice giants of the Solar System. In-situ formation that would involve a large amount of H/He remains difficult to consider, and a formation within or past the snow line of their protoplanetary disk, followed by inward migration appears to provide a more robust alternative. If their formation took place beyond the ice line, this numerous Neptune-class population might have experienced similar formation conditions than our ice giants. This makes even more crucial to get clear references to which compare our exoplanets.

Still today, Neptune and Uranus remain however poorly known. Their atmospheric composition is unfortunately not well known and nor is their internal composition. Their frozen atmosphere might be representative of the early solar system conditions beyond the ice line, depending on the orbital and internal evolution of these planets. Also, a clear determination of the composition of their core (ice-rich or rocky) would be very valuable in the framework of the super-Earths vs. exo-Neptunes dichotomy resulting from efficient photo-evaporation.  

Progress towards a better determination of the bulk and atmospheric compositions of Neptune-class planets and solar system ice giants, allowing their direct comparison, is expected in the coming decades from the emergence of new, more powerful observation tools: on the exoplanet side, JWST and the next generation of giant telescopes. For Uranus and Neptune, an ambitious  program of space exploration of ice giants combining in situ measurements in their atmospheres by atmospheric entry probes with complementary determinations of their gravity and magnetic fields, and therefore of their internal structure. 

Given the abundance and diversity of the exo-Neptune population, it is possible we will gain insights into our solar system’s ice giants placing our planets in a broader evolutionary context.

\bibliographystyle{aps-nameyear}      
\bibliography{IceG.bib}                

\begin{thebibliography}{86}
\ifx \bisbn   \undefined \def \bisbn  #1{ISBN #1}\fi
\ifx \binits  \undefined \def \binits#1{#1} \fi
\ifx \bauthor  \undefined \def \bauthor#1{#1} \fi
\ifx \bjtitle  \undefined \def \bjtitle#1{\textrm{#1}}\fi
\ifx \batitle  \undefined \def \batitle#1{#1} \fi
\ifx \bctitle  \undefined \def \bctitle#1{#1} \fi
\ifx \bvolume  \undefined \def \bvolume#1{\textbf{#1}}\fi
\ifx \byear  \undefined \def \byear#1{#1} \fi
\ifx \bissue  \undefined \def \bissue#1{#1} \fi
\ifx \bfpage  \undefined \def \bfpage#1{#1} \fi
\ifx \blpage  \undefined \def \blpage #1{#1} \fi
\ifx \burl  \undefined \def \burl#1{#1} \fi
\ifx \doiurl  \undefined \def \doiurl#1{#1} \fi
\ifx \betal  \undefined \def \betal{et al.} \fi
\ifx \binstitute  \undefined \def \binstitute#1{#1} \fi
\ifx \beditor  \undefined \def \beditor#1{#1} \fi
\ifx \bpublisher  \undefined \def \bpublisher#1{#1} \fi
\ifx \bbtitle  \undefined \def \bbtitle#1{\textit{#1}} \fi
\ifx \bedition  \undefined \def \bedition#1{#1} \fi
\ifx \bseriesno  \undefined \def \bseriesno#1{#1} \fi
\ifx \blocation  \undefined \def \blocation#1{#1} \fi
\ifx \bsertitle  \undefined \def \bsertitle#1{#1} \fi
\ifx \bsnm \undefined \def \bsnm#1{#1} \fi
\ifx \bsuffix \undefined \def \bsuffix#1{#1} \fi
\ifx \bparticle \undefined \def \bparticle#1{#1} \fi
\ifx \barticle \undefined \def \barticle#1{#1} \fi
\ifx \botherref \undefined \def \botherref #1{#1} \fi
\ifx \url \undefined \def \url#1{#1} \fi
\ifx \bchapter \undefined \def \bchapter#1{#1} \fi
\ifx \bbook \undefined \def \bbook#1{#1} \fi
\ifx \bcomment \undefined \def \bcomment#1{#1} \fi
\ifx \oauthor \undefined \def \oauthor#1{#1} \fi
\ifx \citeauthoryear \undefined \def \citeauthoryear#1{#1} \fi
\ifx \texttildelow  \undefined \def \texttildelow{\symbol{126}} \fi
\def \endbibitem {}
\ifx \bconflocation  \undefined \def \bconflocation#1{#1} \fi

\bibitem[\protect\citeauthoryear{{Allart} et~al.}{2018}]{Allart2018}
\begin{barticle}
\bauthor{\binits{R.} \bsnm{{Allart}}},
\bauthor{\binits{V.} \bsnm{{Bourrier}}},
\bauthor{\binits{C.} \bsnm{{Lovis}}},
\bauthor{\binits{D.} \bsnm{{Ehrenreich}}},
\bauthor{\binits{J.J.} \bsnm{{Spake}}},
\bauthor{\binits{A.} \bsnm{{Wyttenbach}}},
\bauthor{\binits{L.} \bsnm{{Pino}}},
\bauthor{\binits{F.} \bsnm{{Pepe}}},
\bauthor{\binits{D.K.} \bsnm{{Sing}}},
\bauthor{\binits{A.} \bsnm{{Lecavelier des Etangs}}},
\batitle{{Spectrally resolved helium absorption from the extended atmosphere of
  a warm Neptune-mass exoplanet}}.
\bjtitle{Science}
\bvolume{362}(\bissue{6421}),
\bfpage{1384}--\blpage{1387}
(\byear{2018}).
doi:\doiurl{10.1126/science.aat5879}
\end{barticle}
\endbibitem

\bibitem[\protect\citeauthoryear{Armstrong et~al.}{2020}]{Armstrong2020}
\begin{barticle}
\bauthor{\binits{D.J.} \bsnm{Armstrong}},
\bauthor{\binits{T.A.} \bsnm{Lopez}},
\bauthor{\binits{V.} \bsnm{Adibekyan}},
\bauthor{\binits{R.A.} \bsnm{Booth}},
\bauthor{\binits{E.M.} \bsnm{Bryant}},
\bauthor{\binits{K.A.} \bsnm{Collins}},
\bauthor{\binits{M.} \bsnm{Deleuil}},
\bauthor{\binits{A.} \bsnm{Emsenhuber}},
\bauthor{\binits{C.X.} \bsnm{Huang}},
\bauthor{\binits{G.W.} \bsnm{King}}, \betal,
\batitle{A remnant planetary core in the hot-neptune desert}.
\bjtitle{Nature}
\bvolume{583}(\bissue{7814}),
\bfpage{39}--\blpage{42}
(\byear{2020})
\end{barticle}
\endbibitem

\bibitem[\protect\citeauthoryear{{Atreya} et~al.}{2020}]{Atreya2020}
\begin{barticle}
\bauthor{\binits{S.K.} \bsnm{{Atreya}}},
\bauthor{\binits{M.H.} \bsnm{{Hofstadter}}},
\bauthor{\binits{J.H.} \bsnm{{In}}},
\bauthor{\binits{O.} \bsnm{{Mousis}}},
\bauthor{\binits{K.} \bsnm{{Reh}}},
\bauthor{\binits{M.H.} \bsnm{{Wong}}},
\batitle{{Deep Atmosphere Composition, Structure, Origin, and Exploration, with
  Particular Focus on Critical in situ Science at the Icy Giants}}.
\bjtitle{\ssr}
\bvolume{216}(\bissue{1}),
\bfpage{18}
(\byear{2020}).
doi:\doiurl{10.1007/s11214-020-0640-8}
\end{barticle}
\endbibitem

\bibitem[\protect\citeauthoryear{{Barros} et~al.}{2015}]{Barros2015}
\begin{barticle}
\bauthor{\binits{S.C.C.} \bsnm{{Barros}}},
\bauthor{\binits{J.M.} \bsnm{{Almenara}}},
\bauthor{\binits{O.} \bsnm{{Demangeon}}},
\bauthor{\binits{M.} \bsnm{{Tsantaki}}},
\bauthor{\binits{A.} \bsnm{{Santerne}}},
\bauthor{\binits{D.J.} \bsnm{{Armstrong}}},
\bauthor{\binits{D.} \bsnm{{Barrado}}},
\bauthor{\binits{D.} \bsnm{{Brown}}},
\bauthor{\binits{M.} \bsnm{{Deleuil}}},
\bauthor{\binits{J.} \bsnm{{Lillo-Box}}},
\bauthor{\binits{H.} \bsnm{{Osborn}}},
\bauthor{\binits{D.} \bsnm{{Pollacco}}},
\bauthor{\binits{L.} \bsnm{{Abe}}},
\bauthor{\binits{P.} \bsnm{{Andre}}},
\bauthor{\binits{P.} \bsnm{{Bendjoya}}},
\bauthor{\binits{I.} \bsnm{{Boisse}}},
\bauthor{\binits{A.S.} \bsnm{{Bonomo}}},
\bauthor{\binits{F.} \bsnm{{Bouchy}}},
\bauthor{\binits{G.} \bsnm{{Bruno}}},
\bauthor{\binits{J.R.} \bsnm{{Cerda}}},
\bauthor{\binits{B.} \bsnm{{Courcol}}},
\bauthor{\binits{R.F.} \bsnm{{D{\'\i}az}}},
\bauthor{\binits{G.} \bsnm{{H{\'e}brard}}},
\bauthor{\binits{J.} \bsnm{{Kirk}}},
\bauthor{\binits{J.C.} \bsnm{{Lachuri{\'e}}}},
\bauthor{\binits{K.W.F.} \bsnm{{Lam}}},
\bauthor{\binits{P.} \bsnm{{Martinez}}},
\bauthor{\binits{J.} \bsnm{{McCormac}}},
\bauthor{\binits{C.} \bsnm{{Moutou}}},
\bauthor{\binits{A.} \bsnm{{Rajpurohit}}},
\bauthor{\binits{J.-P.} \bsnm{{Rivet}}},
\bauthor{\binits{J.} \bsnm{{Spake}}},
\bauthor{\binits{O.} \bsnm{{Suarez}}},
\bauthor{\binits{D.} \bsnm{{Toublanc}}},
\bauthor{\binits{S.R.} \bsnm{{Walker}}},
\batitle{{Photodynamical mass determination of the multiplanetary system
  K2-19}}.
\bjtitle{\mnras}
\bvolume{454}(\bissue{4}),
\bfpage{4267}--\blpage{4276}
(\byear{2015}).
doi:\doiurl{10.1093/mnras/stv2271}
\end{barticle}
\endbibitem

\bibitem[\protect\citeauthoryear{{Baruteau} and
  {Papaloizou}}{2013}]{Baruteau13}
\begin{barticle}
\bauthor{\binits{C.} \bsnm{{Baruteau}}},
\bauthor{\binits{J.C.B.} \bsnm{{Papaloizou}}},
\batitle{{Disk-Planets Interactions and the Diversity of Period Ratios in
  Kepler's Multi-planetary Systems}}.
\bjtitle{\apj}
\bvolume{778},
\bfpage{7}
(\byear{2013}).
doi:\doiurl{10.1088/0004-637X/778/1/7}
\end{barticle}
\endbibitem

\bibitem[\protect\citeauthoryear{{Benneke} et~al.}{2019}]{Benneke2019}
\begin{barticle}
\bauthor{\binits{B.} \bsnm{{Benneke}}},
\bauthor{\binits{I.} \bsnm{{Wong}}},
\bauthor{\binits{C.} \bsnm{{Piaulet}}},
\bauthor{\binits{H.A.} \bsnm{{Knutson}}},
\bauthor{\binits{J.} \bsnm{{Lothringer}}},
\bauthor{\binits{C.V.} \bsnm{{Morley}}},
\bauthor{\binits{I.J.M.} \bsnm{{Crossfield}}},
\bauthor{\binits{P.} \bsnm{{Gao}}},
\bauthor{\binits{T.P.} \bsnm{{Greene}}},
\bauthor{\binits{C.} \bsnm{{Dressing}}},
\bauthor{\binits{D.} \bsnm{{Dragomir}}},
\bauthor{\binits{A.W.} \bsnm{{Howard}}},
\bauthor{\binits{P.R.} \bsnm{{McCullough}}},
\bauthor{\binits{E.M.-R.} \bsnm{{Kempton}}},
\bauthor{\binits{J.J.} \bsnm{{Fortney}}},
\bauthor{\binits{J.} \bsnm{{Fraine}}},
\batitle{{Water Vapor and Clouds on the Habitable-zone Sub-Neptune Exoplanet
  K2-18b}}.
\bjtitle{\apjl}
\bvolume{887}(\bissue{1}),
\bfpage{14}
(\byear{2019}).
doi:\doiurl{10.3847/2041-8213/ab59dc}
\end{barticle}
\endbibitem

\bibitem[\protect\citeauthoryear{{Benz} et~al.}{2014}]{Benz14}
\begin{bchapter}
\bauthor{\binits{W.} \bsnm{{Benz}}},
\bauthor{\binits{S.} \bsnm{{Ida}}},
\bauthor{\binits{Y.} \bsnm{{Alibert}}},
\bauthor{\binits{D.} \bsnm{{Lin}}},
\bauthor{\binits{C.} \bsnm{{Mordasini}}},
\bctitle{{Planet Population Synthesis}},
in \bbtitle{Protostars and Planets VI},
ed. by \beditor{\binits{H.} \bsnm{{Beuther}}},
\beditor{\binits{R.S.} \bsnm{{Klessen}}},
\beditor{\binits{C.P.} \bsnm{{Dullemond}}},
\beditor{\binits{T.} \bsnm{{Henning}}},
\byear{2014},
pp. \bfpage{691}--\blpage{713}.
doi:\doiurl{10.2458/azu\_uapress\_9780816531240-ch030}
\end{bchapter}
\endbibitem

\bibitem[\protect\citeauthoryear{{Bitsch} et~al.}{2015}]{Bitsch15}
\begin{barticle}
\bauthor{\binits{B.} \bsnm{{Bitsch}}},
\bauthor{\binits{M.} \bsnm{{Lambrechts}}},
\bauthor{\binits{A.} \bsnm{{Johansen}}},
\batitle{{The growth of planets by pebble accretion in evolving protoplanetary
  discs}}.
\bjtitle{\aap}
\bvolume{582},
\bfpage{112}
(\byear{2015}).
doi:\doiurl{10.1051/0004-6361/201526463}
\end{barticle}
\endbibitem

\bibitem[\protect\citeauthoryear{{Blanc} et~al.}{2021}]{Blanc2021}
\begin{barticle}
\bauthor{\binits{M.} \bsnm{{Blanc}}},
\bauthor{\binits{K.} \bsnm{{Mandt}}},
\bauthor{\binits{O.} \bsnm{{Mousis}}},
\bauthor{\binits{N.} \bsnm{{Andr{\'e}}}},
\bauthor{\binits{A.} \bsnm{{Bouquet}}},
\bauthor{\binits{S.} \bsnm{{Charnoz}}},
\bauthor{\binits{K.L.} \bsnm{{Craft}}},
\bauthor{\binits{M.} \bsnm{{Deleuil}}},
\bauthor{\binits{L.} \bsnm{{Griton}}},
\bauthor{\binits{R.} \bsnm{{Helled}}},
\bauthor{\binits{R.} \bsnm{{Hueso}}},
\bauthor{\binits{L.} \bsnm{{Lamy}}},
\bauthor{\binits{C.} \bsnm{{Louis}}},
\bauthor{\binits{J.} \bsnm{{Lunine}}},
\bauthor{\binits{T.} \bsnm{{Ronnet}}},
\bauthor{\binits{J.} \bsnm{{Schmidt}}},
\bauthor{\binits{K.} \bsnm{{Soderlund}}},
\bauthor{\binits{D.} \bsnm{{Turrini}}},
\bauthor{\binits{E.} \bsnm{{Turtle}}},
\bauthor{\binits{P.} \bsnm{{Vernazza}}},
\bauthor{\binits{O.} \bsnm{{Witasse}}},
\batitle{{Science Goals and Mission Objectives for the Future Exploration of
  Ice Giants Systems: A Horizon 2061 Perspective}}.
\bjtitle{\ssr}
\bvolume{217}(\bissue{1}),
\bfpage{3}
(\byear{2021}).
doi:\doiurl{10.1007/s11214-020-00769-5}
\end{barticle}
\endbibitem

\bibitem[\protect\citeauthoryear{{Boehle} et~al.}{2019}]{Boehle2019}
\begin{botherref}
\oauthor{\binits{A.} \bsnm{{Boehle}}},
\oauthor{\binits{S.P.} \bsnm{{Quanz}}},
\oauthor{\binits{C.} \bsnm{{Lovis}}},
\oauthor{\binits{D.} \bsnm{{S{\`e}gransan}}},
\oauthor{\binits{S.} \bsnm{{Udry}}},
\oauthor{\binits{D.} \bsnm{{Apai}}},
{Combining high contrast imaging and radial velocities to constrain the
  planetary architecture of nearby stars}.
arXiv e-prints,
1907--04334
(2019)
\end{botherref}
\endbibitem

\bibitem[\protect\citeauthoryear{{Boisse} et~al.}{2012}]{Boisse2012}
\begin{barticle}
\bauthor{\binits{I.} \bsnm{{Boisse}}},
\bauthor{\binits{F.} \bsnm{{Pepe}}},
\bauthor{\binits{C.} \bsnm{{Perrier}}},
\bauthor{\binits{D.} \bsnm{{Queloz}}},
\bauthor{\binits{X.} \bsnm{{Bonfils}}},
\bauthor{\binits{F.} \bsnm{{Bouchy}}},
\bauthor{\binits{N.C.} \bsnm{{Santos}}},
\bauthor{\binits{L.} \bsnm{{Arnold}}},
\bauthor{\binits{J.-L.} \bsnm{{Beuzit}}},
\bauthor{\binits{R.F.} \bsnm{{D{\'\i}az}}},
\bauthor{\binits{X.} \bsnm{{Delfosse}}},
\bauthor{\binits{A.} \bsnm{{Eggenberger}}},
\bauthor{\binits{D.} \bsnm{{Ehrenreich}}},
\bauthor{\binits{T.} \bsnm{{Forveille}}},
\bauthor{\binits{G.} \bsnm{{H{\'e}brard}}},
\bauthor{\binits{A.-M.} \bsnm{{Lagrange}}},
\bauthor{\binits{C.} \bsnm{{Lovis}}},
\bauthor{\binits{M.} \bsnm{{Mayor}}},
\bauthor{\binits{C.} \bsnm{{Moutou}}},
\bauthor{\binits{D.} \bsnm{{Naef}}},
\bauthor{\binits{A.} \bsnm{{Santerne}}},
\bauthor{\binits{D.} \bsnm{{S{\'e}gransan}}},
\bauthor{\binits{J.-P.} \bsnm{{Sivan}}},
\bauthor{\binits{S.} \bsnm{{Udry}}},
\batitle{{The SOPHIE search for northern extrasolar planets. V. Follow-up of
  ELODIE candidates: Jupiter-analogs around Sun-like stars}}.
\bjtitle{\aap}
\bvolume{545},
\bfpage{55}
(\byear{2012}).
doi:\doiurl{10.1051/0004-6361/201118419}
\end{barticle}
\endbibitem

\bibitem[\protect\citeauthoryear{{Bolton} et~al.}{2017}]{Bolton2017}
\begin{barticle}
\bauthor{\binits{S.J.} \bsnm{{Bolton}}},
\bauthor{\binits{J.} \bsnm{{Lunine}}},
\bauthor{\binits{D.} \bsnm{{Stevenson}}},
\bauthor{\binits{J.E.P.} \bsnm{{Connerney}}},
\bauthor{\binits{S.} \bsnm{{Levin}}},
\bauthor{\binits{T.C.} \bsnm{{Owen}}},
\bauthor{\binits{F.} \bsnm{{Bagenal}}},
\bauthor{\binits{D.} \bsnm{{Gautier}}},
\bauthor{\binits{A.P.} \bsnm{{Ingersoll}}},
\bauthor{\binits{G.S.} \bsnm{{Orton}}},
\bauthor{\binits{T.} \bsnm{{Guillot}}},
\bauthor{\binits{W.} \bsnm{{Hubbard}}},
\bauthor{\binits{J.} \bsnm{{Bloxham}}},
\bauthor{\binits{A.} \bsnm{{Coradini}}},
\bauthor{\binits{S.K.} \bsnm{{Stephens}}},
\bauthor{\binits{P.} \bsnm{{Mokashi}}},
\bauthor{\binits{R.} \bsnm{{Thorne}}},
\bauthor{\binits{R.} \bsnm{{Thorpe}}},
\batitle{{The Juno Mission}}.
\bjtitle{\ssr}
\bvolume{213}(\bissue{1-4}),
\bfpage{5}--\blpage{37}
(\byear{2017}).
doi:\doiurl{10.1007/s11214-017-0429-6}
\end{barticle}
\endbibitem

\bibitem[\protect\citeauthoryear{{Bourrier} et~al.}{2018}]{Bourrier2018}
\begin{barticle}
\bauthor{\binits{V.} \bsnm{{Bourrier}}},
\bauthor{\binits{A.} \bsnm{{Lecavelier des Etangs}}},
\bauthor{\binits{D.} \bsnm{{Ehrenreich}}},
\bauthor{\binits{J.} \bsnm{{Sanz-Forcada}}},
\bauthor{\binits{R.} \bsnm{{Allart}}},
\bauthor{\binits{G.E.} \bsnm{{Ballester}}},
\bauthor{\binits{L.A.} \bsnm{{Buchhave}}},
\bauthor{\binits{O.} \bsnm{{Cohen}}},
\bauthor{\binits{D.} \bsnm{{Deming}}},
\bauthor{\binits{T.M.} \bsnm{{Evans}}},
\bauthor{\binits{A.} \bsnm{{Garc{\'\i}a Mu{\~n}oz}}},
\bauthor{\binits{G.W.} \bsnm{{Henry}}},
\bauthor{\binits{T.} \bsnm{{Kataria}}},
\bauthor{\binits{P.} \bsnm{{Lavvas}}},
\bauthor{\binits{N.} \bsnm{{Lewis}}},
\bauthor{\binits{M.} \bsnm{{L{\'o}pez-Morales}}},
\bauthor{\binits{M.} \bsnm{{Marley}}},
\bauthor{\binits{D.K.} \bsnm{{Sing}}},
\bauthor{\binits{H.R.} \bsnm{{Wakeford}}},
\batitle{{Hubble PanCET: an extended upper atmosphere of neutral hydrogen
  around the warm Neptune GJ 3470b}}.
\bjtitle{\aap}
\bvolume{620},
\bfpage{147}
(\byear{2018}).
doi:\doiurl{10.1051/0004-6361/201833675}
\end{barticle}
\endbibitem

\bibitem[\protect\citeauthoryear{Bourrier et~al.}{2015}]{Bourrier2015}
\begin{barticle}
\bauthor{\binits{V.} \bsnm{Bourrier}},
\bauthor{\binits{D.} \bsnm{Ehrenreich}},
\bauthor{\binits{A.L.} \bparticle{des} \bsnm{Etangs}},
\batitle{Radiative braking in the extended exosphere of gj 436 b}.
\bjtitle{Astronomy \& Astrophysics}
\bvolume{582},
\bfpage{65}
(\byear{2015})
\end{barticle}
\endbibitem

\bibitem[\protect\citeauthoryear{{Bryan} et~al.}{2019}]{Bryan2019}
\begin{barticle}
\bauthor{\binits{M.L.} \bsnm{{Bryan}}},
\bauthor{\binits{H.A.} \bsnm{{Knutson}}},
\bauthor{\binits{E.J.} \bsnm{{Lee}}},
\bauthor{\binits{B.J.} \bsnm{{Fulton}}},
\bauthor{\binits{K.} \bsnm{{Batygin}}},
\bauthor{\binits{H.} \bsnm{{Ngo}}},
\bauthor{\binits{T.} \bsnm{{Meshkat}}},
\batitle{{An Excess of Jupiter Analogs in Super-Earth Systems}}.
\bjtitle{\aj}
\bvolume{157}(\bissue{2}),
\bfpage{52}
(\byear{2019}).
doi:\doiurl{10.3847/1538-3881/aaf57f}
\end{barticle}
\endbibitem

\bibitem[\protect\citeauthoryear{Caldas et~al.}{2019}]{Caldas2018}
\begin{barticle}
\bauthor{\binits{A.} \bsnm{Caldas}},
\bauthor{\binits{J.} \bsnm{Leconte}},
\bauthor{\binits{F.} \bsnm{Selsis}},
\bauthor{\binits{I.} \bsnm{Waldmann}},
\bauthor{\binits{P.} \bsnm{Bord{\'e}}},
\bauthor{\binits{M.} \bsnm{Rocchetto}},
\bauthor{\binits{B.} \bsnm{Charnay}},
\batitle{Effects of a fully 3d atmospheric structure on exoplanet transmission
  spectra: retrieval biases due to day--night temperature gradients}.
\bjtitle{Astronomy \& Astrophysics}
\bvolume{623},
\bfpage{161}
(\byear{2019})
\end{barticle}
\endbibitem

\bibitem[\protect\citeauthoryear{{Childs} et~al.}{2019}]{Childs2019}
\begin{barticle}
\bauthor{\binits{A.C.} \bsnm{{Childs}}},
\bauthor{\binits{E.} \bsnm{{Quintana}}},
\bauthor{\binits{T.} \bsnm{{Barclay}}},
\bauthor{\binits{J.H.} \bsnm{{Steffen}}},
\batitle{{Giant planet effects on terrestrial planet formation and system
  architecture}}.
\bjtitle{\mnras}
\bvolume{485}(\bissue{1}),
\bfpage{541}--\blpage{549}
(\byear{2019}).
doi:\doiurl{10.1093/mnras/stz385}
\end{barticle}
\endbibitem

\bibitem[\protect\citeauthoryear{{Courcol} et~al.}{2016}]{Courcol2016}
\begin{barticle}
\bauthor{\binits{B.} \bsnm{{Courcol}}},
\bauthor{\binits{F.} \bsnm{{Bouchy}}},
\bauthor{\binits{M.} \bsnm{{Deleuil}}},
\batitle{{An upper boundary in the mass-metallicity plane of exo-Neptunes}}.
\bjtitle{\mnras}
\bvolume{461}(\bissue{2}),
\bfpage{1841}--\blpage{1849}
(\byear{2016}).
doi:\doiurl{10.1093/mnras/stw1049}
\end{barticle}
\endbibitem

\bibitem[\protect\citeauthoryear{{dos Santos} et~al.}{2019}]{dosSantos2019}
\begin{barticle}
\bauthor{\binits{L.A.} \bsnm{{dos Santos}}},
\bauthor{\binits{D.} \bsnm{{Ehrenreich}}},
\bauthor{\binits{V.} \bsnm{{Bourrier}}},
\bauthor{\binits{A.} \bsnm{{Lecavelier des Etangs}}},
\bauthor{\binits{M.} \bsnm{{L{\'o}pez-Morales}}},
\bauthor{\binits{D.K.} \bsnm{{Sing}}},
\bauthor{\binits{G.} \bsnm{{Ballester}}},
\bauthor{\binits{L.} \bsnm{{Ben-Jaffel}}},
\bauthor{\binits{L.A.} \bsnm{{Buchhave}}},
\bauthor{\binits{A.} \bsnm{{Garc{\'\i}a Mu{\~n}oz}}},
\bauthor{\binits{G.W.} \bsnm{{Henry}}},
\bauthor{\binits{T.} \bsnm{{Kataria}}},
\bauthor{\binits{B.} \bsnm{{Lavie}}},
\bauthor{\binits{P.} \bsnm{{Lavvas}}},
\bauthor{\binits{N.K.} \bsnm{{Lewis}}},
\bauthor{\binits{T.} \bsnm{{Mikal-Evans}}},
\bauthor{\binits{J.} \bsnm{{Sanz-Forcada}}},
\bauthor{\binits{H.} \bsnm{{Wakeford}}},
\batitle{{The Hubble PanCET program: an extensive search for metallic ions in
  the exosphere of GJ 436 b}}.
\bjtitle{\aap}
\bvolume{629},
\bfpage{47}
(\byear{2019}).
doi:\doiurl{10.1051/0004-6361/201935663}
\end{barticle}
\endbibitem

\bibitem[\protect\citeauthoryear{{dos Santos} et~al.}{2020}]{dosSantos2020}
\begin{barticle}
\bauthor{\binits{L.A.} \bsnm{{dos Santos}}},
\bauthor{\binits{D.} \bsnm{{Ehrenreich}}},
\bauthor{\binits{V.} \bsnm{{Bourrier}}},
\bauthor{\binits{N.} \bsnm{{Astudillo-Defru}}},
\bauthor{\binits{X.} \bsnm{{Bonfils}}},
\bauthor{\binits{F.} \bsnm{{Forget}}},
\bauthor{\binits{C.} \bsnm{{Lovis}}},
\bauthor{\binits{F.} \bsnm{{Pepe}}},
\bauthor{\binits{S.} \bsnm{{Udry}}},
\batitle{{The high-energy environment and atmospheric escape of the
  mini-Neptune K2-18 b}}.
\bjtitle{\aap}
\bvolume{634},
\bfpage{4}
(\byear{2020}).
doi:\doiurl{10.1051/0004-6361/201937327}
\end{barticle}
\endbibitem

\bibitem[\protect\citeauthoryear{{Ehrenreich} et~al.}{2015}]{Ehrenreich2015}
\begin{barticle}
\bauthor{\binits{D.} \bsnm{{Ehrenreich}}},
\bauthor{\binits{V.} \bsnm{{Bourrier}}},
\bauthor{\binits{P.J.} \bsnm{{Wheatley}}},
\bauthor{\binits{A.} \bsnm{{Lecavelier des Etangs}}},
\bauthor{\binits{G.} \bsnm{{H{\'e}brard}}},
\bauthor{\binits{S.} \bsnm{{Udry}}},
\bauthor{\binits{X.} \bsnm{{Bonfils}}},
\bauthor{\binits{X.} \bsnm{{Delfosse}}},
\bauthor{\binits{J.-M.} \bsnm{{D{\'e}sert}}},
\bauthor{\binits{D.K.} \bsnm{{Sing}}},
\bauthor{\binits{A.} \bsnm{{Vidal-Madjar}}},
\batitle{{A giant comet-like cloud of hydrogen escaping the warm Neptune-mass
  exoplanet GJ 436b}}.
\bjtitle{\nat}
\bvolume{522}(\bissue{7557}),
\bfpage{459}--\blpage{461}
(\byear{2015}).
doi:\doiurl{10.1038/nature14501}
\end{barticle}
\endbibitem

\bibitem[\protect\citeauthoryear{{Fabrycky} et~al.}{2014}]{Fabrycky2014}
\begin{barticle}
\bauthor{\binits{D.C.} \bsnm{{Fabrycky}}},
\bauthor{\binits{J.J.} \bsnm{{Lissauer}}},
\bauthor{\binits{D.} \bsnm{{Ragozzine}}},
\bauthor{\binits{J.F.} \bsnm{{Rowe}}},
\bauthor{\binits{J.H.} \bsnm{{Steffen}}},
\bauthor{\binits{E.} \bsnm{{Agol}}},
\bauthor{\binits{T.} \bsnm{{Barclay}}},
\bauthor{\binits{N.} \bsnm{{Batalha}}},
\bauthor{\binits{W.} \bsnm{{Borucki}}},
\bauthor{\binits{D.R.} \bsnm{{Ciardi}}},
\bauthor{\binits{E.B.} \bsnm{{Ford}}},
\bauthor{\binits{T.N.} \bsnm{{Gautier}}},
\bauthor{\binits{J.C.} \bsnm{{Geary}}},
\bauthor{\binits{M.J.} \bsnm{{Holman}}},
\bauthor{\binits{J.M.} \bsnm{{Jenkins}}},
\bauthor{\binits{J.} \bsnm{{Li}}},
\bauthor{\binits{R.C.} \bsnm{{Morehead}}},
\bauthor{\binits{R.L.} \bsnm{{Morris}}},
\bauthor{\binits{A.} \bsnm{{Shporer}}},
\bauthor{\binits{J.C.} \bsnm{{Smith}}},
\bauthor{\binits{M.} \bsnm{{Still}}},
\bauthor{\binits{J.} \bsnm{{Van Cleve}}},
\batitle{{Architecture of Kepler's Multi-transiting Systems. II. New
  Investigations with Twice as Many Candidates}}.
\bjtitle{\apj}
\bvolume{790}(\bissue{2}),
\bfpage{146}
(\byear{2014}).
doi:\doiurl{10.1088/0004-637X/790/2/146}
\end{barticle}
\endbibitem

\bibitem[\protect\citeauthoryear{{Fulton} and {Petigura}}{2018}]{Fulton2018}
\begin{barticle}
\bauthor{\binits{B.J.} \bsnm{{Fulton}}},
\bauthor{\binits{E.A.} \bsnm{{Petigura}}},
\batitle{{The California-Kepler Survey. VII. Precise Planet Radii Leveraging
  Gaia DR2 Reveal the Stellar Mass Dependence of the Planet Radius Gap}}.
\bjtitle{\aj}
\bvolume{156}(\bissue{6}),
\bfpage{264}
(\byear{2018}).
doi:\doiurl{10.3847/1538-3881/aae828}
\end{barticle}
\endbibitem

\bibitem[\protect\citeauthoryear{{Fulton} et~al.}{2017}]{Fulton2017}
\begin{barticle}
\bauthor{\binits{B.J.} \bsnm{{Fulton}}},
\bauthor{\binits{E.A.} \bsnm{{Petigura}}},
\bauthor{\binits{A.W.} \bsnm{{Howard}}},
\bauthor{\binits{H.} \bsnm{{Isaacson}}},
\bauthor{\binits{G.W.} \bsnm{{Marcy}}},
\bauthor{\binits{P.A.} \bsnm{{Cargile}}},
\bauthor{\binits{L.} \bsnm{{Hebb}}},
\bauthor{\binits{L.M.} \bsnm{{Weiss}}},
\bauthor{\binits{J.A.} \bsnm{{Johnson}}},
\bauthor{\binits{T.D.} \bsnm{{Morton}}},
\bauthor{\binits{E.} \bsnm{{Sinukoff}}},
\bauthor{\binits{I.J.M.} \bsnm{{Crossfield}}},
\bauthor{\binits{L.A.} \bsnm{{Hirsch}}},
\batitle{{The California-Kepler Survey. III. A Gap in the Radius Distribution
  of Small Planets}}.
\bjtitle{\aj}
\bvolume{154}(\bissue{3}),
\bfpage{109}
(\byear{2017}).
doi:\doiurl{10.3847/1538-3881/aa80eb}
\end{barticle}
\endbibitem

\bibitem[\protect\citeauthoryear{{Ginzburg} et~al.}{2018}]{Ginzburg2018}
\begin{barticle}
\bauthor{\binits{S.} \bsnm{{Ginzburg}}},
\bauthor{\binits{H.E.} \bsnm{{Schlichting}}},
\bauthor{\binits{R.} \bsnm{{Sari}}},
\batitle{{Core-powered mass-loss and the radius distribution of small
  exoplanets}}.
\bjtitle{\mnras}
\bvolume{476}(\bissue{1}),
\bfpage{759}--\blpage{765}
(\byear{2018}).
doi:\doiurl{10.1093/mnras/sty290}
\end{barticle}
\endbibitem

\bibitem[\protect\citeauthoryear{{Gupta} and {Schlichting}}{2019}]{Gupta2019}
\begin{barticle}
\bauthor{\binits{A.} \bsnm{{Gupta}}},
\bauthor{\binits{H.E.} \bsnm{{Schlichting}}},
\batitle{{Sculpting the valley in the radius distribution of small exoplanets
  as a by-product of planet formation: the core-powered mass-loss mechanism}}.
\bjtitle{\mnras}
\bvolume{487}(\bissue{1}),
\bfpage{24}--\blpage{33}
(\byear{2019}).
doi:\doiurl{10.1093/mnras/stz1230}
\end{barticle}
\endbibitem

\bibitem[\protect\citeauthoryear{{Gupta} and {Schlichting}}{2020}]{Gupta2020}
\begin{barticle}
\bauthor{\binits{A.} \bsnm{{Gupta}}},
\bauthor{\binits{H.E.} \bsnm{{Schlichting}}},
\batitle{{Signatures of the Core-Powered Mass-Loss Mechanism in the Exoplanet
  Population: Dependence on Stellar Properties and Observational Predictions}}.
\bjtitle{\mnras}
(\byear{2020}).
doi:\doiurl{10.1093/mnras/staa315}
\end{barticle}
\endbibitem

\bibitem[\protect\citeauthoryear{{Hansen} and {Murray}}{2013}]{Hansen13}
\begin{barticle}
\bauthor{\binits{B.M.S.} \bsnm{{Hansen}}},
\bauthor{\binits{N.} \bsnm{{Murray}}},
\batitle{{Testing in Situ Assembly with the Kepler Planet Candidate Sample}}.
\bjtitle{\apj}
\bvolume{775},
\bfpage{53}
(\byear{2013}).
doi:\doiurl{10.1088/0004-637X/775/1/53}
\end{barticle}
\endbibitem

\bibitem[\protect\citeauthoryear{{Izidoro} et~al.}{2015}]{Izidoro15}
\begin{barticle}
\bauthor{\binits{A.} \bsnm{{Izidoro}}},
\bauthor{\binits{A.} \bsnm{{Morbidelli}}},
\bauthor{\binits{S.N.} \bsnm{{Raymond}}},
\bauthor{\binits{F.} \bsnm{{Hersant}}},
\bauthor{\binits{A.} \bsnm{{Pierens}}},
\batitle{{Accretion of Uranus and Neptune from inward-migrating planetary
  embryos blocked by Jupiter and Saturn}}.
\bjtitle{\aap}
\bvolume{582},
\bfpage{99}
(\byear{2015}).
doi:\doiurl{10.1051/0004-6361/201425525}
\end{barticle}
\endbibitem

\bibitem[\protect\citeauthoryear{{Izidoro} et~al.}{2017}]{Izidoro2017}
\begin{barticle}
\bauthor{\binits{A.} \bsnm{{Izidoro}}},
\bauthor{\binits{M.} \bsnm{{Ogihara}}},
\bauthor{\binits{S.N.} \bsnm{{Raymond}}},
\bauthor{\binits{A.} \bsnm{{Morbidelli}}},
\bauthor{\binits{A.} \bsnm{{Pierens}}},
\bauthor{\binits{B.} \bsnm{{Bitsch}}},
\bauthor{\binits{C.} \bsnm{{Cossou}}},
\bauthor{\binits{F.} \bsnm{{Hersant}}},
\batitle{{Breaking the chains: hot super-Earth systems from migration and
  disruption of compact resonant chains}}.
\bjtitle{\mnras}
\bvolume{470}(\bissue{2}),
\bfpage{1750}--\blpage{1770}
(\byear{2017}).
doi:\doiurl{10.1093/mnras/stx1232}
\end{barticle}
\endbibitem

\bibitem[\protect\citeauthoryear{{Jin} and {Mordasini}}{2018}]{Jin2018}
\begin{barticle}
\bauthor{\binits{S.} \bsnm{{Jin}}},
\bauthor{\binits{C.} \bsnm{{Mordasini}}},
\batitle{{Compositional Imprints in Density-Distance-Time: A Rocky Composition
  for Close-in Low-mass Exoplanets from the Location of the Valley of
  Evaporation}}.
\bjtitle{\apj}
\bvolume{853}(\bissue{2}),
\bfpage{163}
(\byear{2018}).
doi:\doiurl{10.3847/1538-4357/aa9f1e}
\end{barticle}
\endbibitem

\bibitem[\protect\citeauthoryear{{Jontof-Hutter}
  et~al.}{2015}]{Jontof-Hutter2015}
\begin{barticle}
\bauthor{\binits{D.} \bsnm{{Jontof-Hutter}}},
\bauthor{\binits{J.F.} \bsnm{{Rowe}}},
\bauthor{\binits{J.J.} \bsnm{{Lissauer}}},
\bauthor{\binits{D.C.} \bsnm{{Fabrycky}}},
\bauthor{\binits{E.B.} \bsnm{{Ford}}},
\batitle{{The mass of the Mars-sized exoplanet Kepler-138 b from transit
  timing}}.
\bjtitle{\nat}
\bvolume{522}(\bissue{7556}),
\bfpage{321}--\blpage{323}
(\byear{2015}).
doi:\doiurl{10.1038/nature14494}
\end{barticle}
\endbibitem

\bibitem[\protect\citeauthoryear{{Kane} et~al.}{2019}]{Kane2019}
\begin{barticle}
\bauthor{\binits{S.R.} \bsnm{{Kane}}},
\bauthor{\binits{P.A.} \bsnm{{Dalba}}},
\bauthor{\binits{Z.} \bsnm{{Li}}},
\bauthor{\binits{E.P.} \bsnm{{Horch}}},
\bauthor{\binits{L.A.} \bsnm{{Hirsch}}},
\bauthor{\binits{J.} \bsnm{{Horner}}},
\bauthor{\binits{R.A.} \bsnm{{Wittenmyer}}},
\bauthor{\binits{S.B.} \bsnm{{Howell}}},
\bauthor{\binits{M.E.} \bsnm{{Everett}}},
\bauthor{\binits{R.P.} \bsnm{{Butler}}},
\bauthor{\binits{C.G.} \bsnm{{Tinney}}},
\bauthor{\binits{B.D.} \bsnm{{Carter}}},
\bauthor{\binits{D.J.} \bsnm{{Wright}}},
\bauthor{\binits{H.R.A.} \bsnm{{Jones}}},
\bauthor{\binits{J.} \bsnm{{Bailey}}},
\bauthor{\binits{S.J.} \bsnm{{O{\textquoteright}Toole}}},
\batitle{{Detection of Planetary and Stellar Companions to Neighboring Stars
  via a Combination of Radial Velocity and Direct Imaging Techniques}}.
\bjtitle{\aj}
\bvolume{157}(\bissue{6}),
\bfpage{252}
(\byear{2019}).
doi:\doiurl{10.3847/1538-3881/ab1ddf}
\end{barticle}
\endbibitem

\bibitem[\protect\citeauthoryear{{Kipping} et~al.}{2016}]{Kipping2016}
\begin{barticle}
\bauthor{\binits{D.M.} \bsnm{{Kipping}}},
\bauthor{\binits{G.} \bsnm{{Torres}}},
\bauthor{\binits{C.} \bsnm{{Henze}}},
\bauthor{\binits{A.} \bsnm{{Teachey}}},
\bauthor{\binits{H.} \bsnm{{Isaacson}}},
\bauthor{\binits{E.} \bsnm{{Petigura}}},
\bauthor{\binits{G.W.} \bsnm{{Marcy}}},
\bauthor{\binits{L.A.} \bsnm{{Buchhave}}},
\bauthor{\binits{J.} \bsnm{{Chen}}},
\bauthor{\binits{S.T.} \bsnm{{Bryson}}},
\bauthor{\binits{E.} \bsnm{{Sandford}}},
\batitle{{A Transiting Jupiter Analog}}.
\bjtitle{\apj}
\bvolume{820}(\bissue{2}),
\bfpage{112}
(\byear{2016}).
doi:\doiurl{10.3847/0004-637X/820/2/112}
\end{barticle}
\endbibitem

\bibitem[\protect\citeauthoryear{{Komacek} et~al.}{2019}]{Komacek2019}
\begin{barticle}
\bauthor{\binits{T.D.} \bsnm{{Komacek}}},
\bauthor{\binits{A.P.} \bsnm{{Showman}}},
\bauthor{\binits{V.} \bsnm{{Parmentier}}},
\batitle{{Vertical Tracer Mixing in Hot Jupiter Atmospheres}}.
\bjtitle{\apj}
\bvolume{881}(\bissue{2}),
\bfpage{152}
(\byear{2019}).
doi:\doiurl{10.3847/1538-4357/ab338b}
\end{barticle}
\endbibitem

\bibitem[\protect\citeauthoryear{{Lambrechts} and
  {Johansen}}{2014}]{Lambrechts14}
\begin{barticle}
\bauthor{\binits{M.} \bsnm{{Lambrechts}}},
\bauthor{\binits{A.} \bsnm{{Johansen}}},
\batitle{{Forming the cores of giant planets from the radial pebble flux in
  protoplanetary discs}}.
\bjtitle{\aap}
\bvolume{572},
\bfpage{107}
(\byear{2014}).
doi:\doiurl{10.1051/0004-6361/201424343}
\end{barticle}
\endbibitem

\bibitem[\protect\citeauthoryear{{Lambrechts} et~al.}{2019}]{Lambrechts19}
\begin{barticle}
\bauthor{\binits{M.} \bsnm{{Lambrechts}}},
\bauthor{\binits{E.} \bsnm{{Lega}}},
\bauthor{\binits{R.P.} \bsnm{{Nelson}}},
\bauthor{\binits{A.} \bsnm{{Crida}}},
\bauthor{\binits{A.} \bsnm{{Morbidelli}}},
\batitle{{Quasi-static contraction during runaway gas accretion onto giant
  planets}}.
\bjtitle{\aap}
\bvolume{630},
\bfpage{82}
(\byear{2019}).
doi:\doiurl{10.1051/0004-6361/201834413}
\end{barticle}
\endbibitem

\bibitem[\protect\citeauthoryear{{Lammer} et~al.}{2003}]{Lammer2003}
\begin{barticle}
\bauthor{\binits{H.} \bsnm{{Lammer}}},
\bauthor{\binits{F.} \bsnm{{Selsis}}},
\bauthor{\binits{I.} \bsnm{{Ribas}}},
\bauthor{\binits{E.F.} \bsnm{{Guinan}}},
\bauthor{\binits{S.J.} \bsnm{{Bauer}}},
\bauthor{\binits{W.W.} \bsnm{{Weiss}}},
\batitle{{Atmospheric Loss of Exoplanets Resulting from Stellar X-Ray and
  Extreme-Ultraviolet Heating}}.
\bjtitle{\apjl}
\bvolume{598}(\bissue{2}),
\bfpage{121}--\blpage{124}
(\byear{2003}).
doi:\doiurl{10.1086/380815}
\end{barticle}
\endbibitem

\bibitem[\protect\citeauthoryear{{Lavie} et~al.}{2017}]{Lavie2017}
\begin{barticle}
\bauthor{\binits{B.} \bsnm{{Lavie}}},
\bauthor{\binits{D.} \bsnm{{Ehrenreich}}},
\bauthor{\binits{V.} \bsnm{{Bourrier}}},
\bauthor{\binits{A.} \bsnm{{Lecavelier des Etangs}}},
\bauthor{\binits{A.} \bsnm{{Vidal-Madjar}}},
\bauthor{\binits{X.} \bsnm{{Delfosse}}},
\bauthor{\binits{A.} \bsnm{{Gracia Berna}}},
\bauthor{\binits{K.} \bsnm{{Heng}}},
\bauthor{\binits{N.} \bsnm{{Thomas}}},
\bauthor{\binits{S.} \bsnm{{Udry}}},
\bauthor{\binits{P.J.} \bsnm{{Wheatley}}},
\batitle{{The long egress of GJ 436b's giant exosphere}}.
\bjtitle{\aap}
\bvolume{605},
\bfpage{7}
(\byear{2017}).
doi:\doiurl{10.1051/0004-6361/201731340}
\end{barticle}
\endbibitem

\bibitem[\protect\citeauthoryear{{Lissauer} et~al.}{2011}]{Lissauer2011}
\begin{barticle}
\bauthor{\binits{J.J.} \bsnm{{Lissauer}}},
\bauthor{\binits{D.} \bsnm{{Ragozzine}}},
\bauthor{\binits{D.C.} \bsnm{{Fabrycky}}},
\bauthor{\binits{J.H.} \bsnm{{Steffen}}},
\bauthor{\binits{E.B.} \bsnm{{Ford}}},
\bauthor{\binits{J.M.} \bsnm{{Jenkins}}},
\bauthor{\binits{A.} \bsnm{{Shporer}}},
\bauthor{\binits{M.J.} \bsnm{{Holman}}},
\bauthor{\binits{J.F.} \bsnm{{Rowe}}},
\bauthor{\binits{E.V.} \bsnm{{Quintana}}},
\bauthor{\binits{N.M.} \bsnm{{Batalha}}},
\bauthor{\binits{W.J.} \bsnm{{Borucki}}},
\bauthor{\binits{S.T.} \bsnm{{Bryson}}},
\bauthor{\binits{D.A.} \bsnm{{Caldwell}}},
\bauthor{\binits{J.A.} \bsnm{{Carter}}},
\bauthor{\binits{D.} \bsnm{{Ciardi}}},
\bauthor{\binits{E.W.} \bsnm{{Dunham}}},
\bauthor{\binits{J.J.} \bsnm{{Fortney}}},
\bauthor{\binits{I.} \bsnm{{Gautier}} \bsuffix{Thomas~N.}},
\bauthor{\binits{S.B.} \bsnm{{Howell}}},
\bauthor{\binits{D.G.} \bsnm{{Koch}}},
\bauthor{\binits{D.W.} \bsnm{{Latham}}},
\bauthor{\binits{G.W.} \bsnm{{Marcy}}},
\bauthor{\binits{R.C.} \bsnm{{Morehead}}},
\bauthor{\binits{D.} \bsnm{{Sasselov}}},
\batitle{{Architecture and Dynamics of Kepler's Candidate Multiple Transiting
  Planet Systems}}.
\bjtitle{\apjs}
\bvolume{197}(\bissue{1}),
\bfpage{8}
(\byear{2011}).
doi:\doiurl{10.1088/0067-0049/197/1/8}
\end{barticle}
\endbibitem

\bibitem[\protect\citeauthoryear{{MacDonald} and
  {Madhusudhan}}{2019}]{MacDo2019}
\begin{barticle}
\bauthor{\binits{R.J.} \bsnm{{MacDonald}}},
\bauthor{\binits{N.} \bsnm{{Madhusudhan}}},
\batitle{{The metal-rich atmosphere of the exo-Neptune HAT-P-26b}}.
\bjtitle{\mnras}
\bvolume{486}(\bissue{1}),
\bfpage{1292}--\blpage{1315}
(\byear{2019}).
doi:\doiurl{10.1093/mnras/stz789}
\end{barticle}
\endbibitem

\bibitem[\protect\citeauthoryear{{Madhusudhan} et~al.}{2016}]{Madhusudhan2016}
\begin{barticle}
\bauthor{\binits{N.} \bsnm{{Madhusudhan}}},
\bauthor{\binits{M.} \bsnm{{Ag{\'u}ndez}}},
\bauthor{\binits{J.I.} \bsnm{{Moses}}},
\bauthor{\binits{Y.} \bsnm{{Hu}}},
\batitle{{Exoplanetary Atmospheres{\textemdash}Chemistry, Formation Conditions,
  and Habitability}}.
\bjtitle{\ssr}
\bvolume{205}(\bissue{1-4}),
\bfpage{285}--\blpage{348}
(\byear{2016}).
doi:\doiurl{10.1007/s11214-016-0254-3}
\end{barticle}
\endbibitem

\bibitem[\protect\citeauthoryear{Madhusudhan et~al.}{2020}]{Madhu2020}
\begin{barticle}
\bauthor{\binits{N.} \bsnm{Madhusudhan}},
\bauthor{\binits{M.C.} \bsnm{Nixon}},
\bauthor{\binits{L.} \bsnm{Welbanks}},
\bauthor{\binits{A.A.} \bsnm{Piette}},
\bauthor{\binits{R.A.} \bsnm{Booth}},
\batitle{The interior and atmosphere of the habitable-zone exoplanet k2-18b}.
\bjtitle{The Astrophysical Journal Letters}
\bvolume{891}(\bissue{1}),
\bfpage{7}
(\byear{2020})
\end{barticle}
\endbibitem

\bibitem[\protect\citeauthoryear{{Mandt} et~al.}{2020}]{Mandt2020}
\begin{barticle}
\bauthor{\binits{K.E.} \bsnm{{Mandt}}},
\bauthor{\binits{O.} \bsnm{{Mousis}}},
\bauthor{\binits{S.} \bsnm{{Treat}}},
\batitle{{Determining the origin of the building blocks of the Ice Giants based
  on analogue measurements from comets}}.
\bjtitle{\mnras}
\bvolume{491}(\bissue{1}),
\bfpage{488}--\blpage{494}
(\byear{2020}).
doi:\doiurl{10.1093/mnras/stz3061}
\end{barticle}
\endbibitem

\bibitem[\protect\citeauthoryear{{Mansfield} et~al.}{2018}]{Mansfield2018}
\begin{barticle}
\bauthor{\binits{M.} \bsnm{{Mansfield}}},
\bauthor{\binits{J.L.} \bsnm{{Bean}}},
\bauthor{\binits{A.} \bsnm{{Oklop{\v{c}}i{\'c}}}},
\bauthor{\binits{L.} \bsnm{{Kreidberg}}},
\bauthor{\binits{J.-M.} \bsnm{{D{\'e}sert}}},
\bauthor{\binits{E.M.-R.} \bsnm{{Kempton}}},
\bauthor{\binits{M.R.} \bsnm{{Line}}},
\bauthor{\binits{J.J.} \bsnm{{Fortney}}},
\bauthor{\binits{G.W.} \bsnm{{Henry}}},
\bauthor{\binits{M.} \bsnm{{Mallonn}}},
\bauthor{\binits{K.B.} \bsnm{{Stevenson}}},
\bauthor{\binits{D.} \bsnm{{Dragomir}}},
\bauthor{\binits{R.} \bsnm{{Allart}}},
\bauthor{\binits{V.} \bsnm{{Bourrier}}},
\batitle{{Detection of Helium in the Atmosphere of the Exo-Neptune HAT-P-11b}}.
\bjtitle{\apjl}
\bvolume{868}(\bissue{2}),
\bfpage{34}
(\byear{2018}).
doi:\doiurl{10.3847/2041-8213/aaf166}
\end{barticle}
\endbibitem

\bibitem[\protect\citeauthoryear{{Morbidelli} et~al.}{2012}]{Morbidelli2012}
\begin{barticle}
\bauthor{\binits{A.} \bsnm{{Morbidelli}}},
\bauthor{\binits{J.I.} \bsnm{{Lunine}}},
\bauthor{\binits{D.P.} \bsnm{{O'Brien}}},
\bauthor{\binits{S.N.} \bsnm{{Raymond}}},
\bauthor{\binits{K.J.} \bsnm{{Walsh}}},
\batitle{{Building Terrestrial Planets}}.
\bjtitle{Annual Review of Earth and Planetary Sciences}
\bvolume{40}(\bissue{1}),
\bfpage{251}--\blpage{275}
(\byear{2012}).
doi:\doiurl{10.1146/annurev-earth-042711-105319}
\end{barticle}
\endbibitem

\bibitem[\protect\citeauthoryear{{Morley} et~al.}{2017}]{Morley2017}
\begin{barticle}
\bauthor{\binits{C.V.} \bsnm{{Morley}}},
\bauthor{\binits{H.} \bsnm{{Knutson}}},
\bauthor{\binits{M.} \bsnm{{Line}}},
\bauthor{\binits{J.J.} \bsnm{{Fortney}}},
\bauthor{\binits{D.} \bsnm{{Thorngren}}},
\bauthor{\binits{M.S.} \bsnm{{Marley}}},
\bauthor{\binits{D.} \bsnm{{Teal}}},
\bauthor{\binits{R.} \bsnm{{Lupu}}},
\batitle{{Forward and Inverse Modeling of the Emission and Transmission
  Spectrum of GJ 436b: Investigating Metal Enrichment, Tidal Heating, and
  Clouds}}.
\bjtitle{\aj}
\bvolume{153}(\bissue{2}),
\bfpage{86}
(\byear{2017}).
doi:\doiurl{10.3847/1538-3881/153/2/86}
\end{barticle}
\endbibitem

\bibitem[\protect\citeauthoryear{{Motalebi} et~al.}{2015}]{Motalebi2015}
\begin{barticle}
\bauthor{\binits{F.} \bsnm{{Motalebi}}},
\bauthor{\binits{S.} \bsnm{{Udry}}},
\bauthor{\binits{M.} \bsnm{{Gillon}}},
\bauthor{\binits{C.} \bsnm{{Lovis}}},
\bauthor{\binits{D.} \bsnm{{S{\'e}gransan}}},
\bauthor{\binits{L.A.} \bsnm{{Buchhave}}},
\bauthor{\binits{B.O.} \bsnm{{Demory}}},
\bauthor{\binits{L.} \bsnm{{Malavolta}}},
\bauthor{\binits{C.D.} \bsnm{{Dressing}}},
\bauthor{\binits{D.} \bsnm{{Sasselov}}},
\bauthor{\binits{K.} \bsnm{{Rice}}},
\bauthor{\binits{D.} \bsnm{{Charbonneau}}},
\bauthor{\binits{A.} \bsnm{{Collier Cameron}}},
\bauthor{\binits{D.} \bsnm{{Latham}}},
\bauthor{\binits{E.} \bsnm{{Molinari}}},
\bauthor{\binits{F.} \bsnm{{Pepe}}},
\bauthor{\binits{L.} \bsnm{{Affer}}},
\bauthor{\binits{A.S.} \bsnm{{Bonomo}}},
\bauthor{\binits{R.} \bsnm{{Cosentino}}},
\bauthor{\binits{X.} \bsnm{{Dumusque}}},
\bauthor{\binits{P.} \bsnm{{Figueira}}},
\bauthor{\binits{A.F.M.} \bsnm{{Fiorenzano}}},
\bauthor{\binits{S.} \bsnm{{Gettel}}},
\bauthor{\binits{A.} \bsnm{{Harutyunyan}}},
\bauthor{\binits{R.D.} \bsnm{{Haywood}}},
\bauthor{\binits{J.} \bsnm{{Johnson}}},
\bauthor{\binits{E.} \bsnm{{Lopez}}},
\bauthor{\binits{M.} \bsnm{{Lopez-Morales}}},
\bauthor{\binits{M.} \bsnm{{Mayor}}},
\bauthor{\binits{G.} \bsnm{{Micela}}},
\bauthor{\binits{A.} \bsnm{{Mortier}}},
\bauthor{\binits{V.} \bsnm{{Nascimbeni}}},
\bauthor{\binits{D.} \bsnm{{Philips}}},
\bauthor{\binits{G.} \bsnm{{Piotto}}},
\bauthor{\binits{D.} \bsnm{{Pollacco}}},
\bauthor{\binits{D.} \bsnm{{Queloz}}},
\bauthor{\binits{A.} \bsnm{{Sozzetti}}},
\bauthor{\binits{A.} \bsnm{{Vand erburg}}},
\bauthor{\binits{C.A.} \bsnm{{Watson}}},
\batitle{{The HARPS-N Rocky Planet Search. I. HD 219134 b: A transiting rocky
  planet in a multi-planet system at 6.5 pc from the Sun}}.
\bjtitle{\aap}
\bvolume{584},
\bfpage{72}
(\byear{2015}).
doi:\doiurl{10.1051/0004-6361/201526822}
\end{barticle}
\endbibitem

\bibitem[\protect\citeauthoryear{{Mousis} et~al.}{2018}]{Mousis2018}
\begin{barticle}
\bauthor{\binits{O.} \bsnm{{Mousis}}},
\bauthor{\binits{D.H.} \bsnm{{Atkinson}}},
\bauthor{\binits{T.} \bsnm{{Cavali{\'e}}}},
\bauthor{\binits{L.N.} \bsnm{{Fletcher}}},
\bauthor{\binits{M.J.} \bsnm{{Amato}}},
\bauthor{\binits{S.} \bsnm{{Aslam}}},
\bauthor{\binits{F.} \bsnm{{Ferri}}},
\bauthor{\binits{J.-B.} \bsnm{{Renard}}},
\bauthor{\binits{T.} \bsnm{{Spilker}}},
\bauthor{\binits{E.} \bsnm{{Venkatapathy}}},
\bauthor{\binits{P.} \bsnm{{Wurz}}},
\bauthor{\binits{K.} \bsnm{{Aplin}}},
\bauthor{\binits{A.} \bsnm{{Coustenis}}},
\bauthor{\binits{M.} \bsnm{{Deleuil}}},
\bauthor{\binits{M.} \bsnm{{Dobrijevic}}},
\bauthor{\binits{T.} \bsnm{{Fouchet}}},
\bauthor{\binits{T.} \bsnm{{Guillot}}},
\bauthor{\binits{P.} \bsnm{{Hartogh}}},
\bauthor{\binits{T.} \bsnm{{Hewagama}}},
\bauthor{\binits{M.D.} \bsnm{{Hofstadter}}},
\bauthor{\binits{V.} \bsnm{{Hue}}},
\bauthor{\binits{R.} \bsnm{{Hueso}}},
\bauthor{\binits{J.-P.} \bsnm{{Lebreton}}},
\bauthor{\binits{E.} \bsnm{{Lellouch}}},
\bauthor{\binits{J.} \bsnm{{Moses}}},
\bauthor{\binits{G.S.} \bsnm{{Orton}}},
\bauthor{\binits{J.C.} \bsnm{{Pearl}}},
\bauthor{\binits{A.} \bsnm{{S{\'a}nchez-Lavega}}},
\bauthor{\binits{A.} \bsnm{{Simon}}},
\bauthor{\binits{O.} \bsnm{{Venot}}},
\bauthor{\binits{J.H.} \bsnm{{Waite}}},
\bauthor{\binits{R.K.} \bsnm{{Achterberg}}},
\bauthor{\binits{S.} \bsnm{{Atreya}}},
\bauthor{\binits{F.} \bsnm{{Billebaud}}},
\bauthor{\binits{M.} \bsnm{{Blanc}}},
\bauthor{\binits{F.} \bsnm{{Borget}}},
\bauthor{\binits{B.} \bsnm{{Brugger}}},
\bauthor{\binits{S.} \bsnm{{Charnoz}}},
\bauthor{\binits{T.} \bsnm{{Chiavassa}}},
\bauthor{\binits{V.} \bsnm{{Cottini}}},
\bauthor{\binits{L.} \bsnm{{d'Hendecourt}}},
\bauthor{\binits{G.} \bsnm{{Danger}}},
\bauthor{\binits{T.} \bsnm{{Encrenaz}}},
\bauthor{\binits{N.J.P.} \bsnm{{Gorius}}},
\bauthor{\binits{L.} \bsnm{{Jorda}}},
\bauthor{\binits{B.} \bsnm{{Marty}}},
\bauthor{\binits{R.} \bsnm{{Moreno}}},
\bauthor{\binits{A.} \bsnm{{Morse}}},
\bauthor{\binits{C.} \bsnm{{Nixon}}},
\bauthor{\binits{K.} \bsnm{{Reh}}},
\bauthor{\binits{T.} \bsnm{{Ronnet}}},
\bauthor{\binits{F.-X.} \bsnm{{Schmider}}},
\bauthor{\binits{S.} \bsnm{{Sheridan}}},
\bauthor{\binits{C.} \bsnm{{Sotin}}},
\bauthor{\binits{P.} \bsnm{{Vernazza}}},
\bauthor{\binits{G.L.} \bsnm{{Villanueva}}},
\batitle{{Scientific rationale for Uranus and Neptune in situ explorations}}.
\bjtitle{\planss}
\bvolume{155},
\bfpage{12}--\blpage{40}
(\byear{2018}).
doi:\doiurl{10.1016/j.pss.2017.10.005}
\end{barticle}
\endbibitem

\bibitem[\protect\citeauthoryear{{Mousis} et~al.}{2020}]{Mousis2020}
\begin{botherref}
\oauthor{\binits{O.} \bsnm{{Mousis}}},
\oauthor{\binits{M.} \bsnm{{Deleuil}}},
\oauthor{\binits{A.} \bsnm{{Aguichine}}},
\oauthor{\binits{E.} \bsnm{{Marcq}}},
\oauthor{\binits{J.} \bsnm{{Naar}}},
\oauthor{\binits{L.} \bsnm{{Acu{\~n}a Aguirre}}},
\oauthor{\binits{B.} \bsnm{{Brugger}}},
\oauthor{\binits{T.} \bsnm{{Goncalves}}},
{Irradiated ocean planets bridge super-Earth and sub-Neptune populations}.
arXiv e-prints,
2002--05243
(2020)
\end{botherref}
\endbibitem

\bibitem[\protect\citeauthoryear{{Nayakshin}}{2017}]{NayakDawes}
\begin{barticle}
\bauthor{\binits{S.} \bsnm{{Nayakshin}}},
\batitle{{Dawes Review 7: The Tidal Downsizing Hypothesis of Planet
  Formation}}.
\bjtitle{\pasa}
\bvolume{34},
\bfpage{002}
(\byear{2017}).
doi:\doiurl{10.1017/pasa.2016.55}
\end{barticle}
\endbibitem

\bibitem[\protect\citeauthoryear{Ninan et~al.}{2019}]{Ninan2019}
\begin{botherref}
\oauthor{\binits{J.P.} \bsnm{Ninan}},
\oauthor{\binits{G.} \bsnm{Stefansson}},
\oauthor{\binits{S.} \bsnm{Mahadevan}},
\oauthor{\binits{C.} \bsnm{Bender}},
\oauthor{\binits{P.} \bsnm{Robertson}},
\oauthor{\binits{L.} \bsnm{Ramsey}},
\oauthor{\binits{R.} \bsnm{Terrien}},
\oauthor{\binits{J.} \bsnm{Wright}},
\oauthor{\binits{S.A.} \bsnm{Diddams}},
\oauthor{\binits{S.} \bsnm{Kanodia}}, et al.,
Detection of he i 10830$\backslash$aa\~{} absorption during the transit of a
  warm neptune around the m-dwarf gj 3470 with the habitable-zone planet
  finder.
arXiv preprint arXiv:1910.02070
(2019)
\end{botherref}
\endbibitem

\bibitem[\protect\citeauthoryear{{Ogihara} et~al.}{2015}]{Ogihara2015}
\begin{barticle}
\bauthor{\binits{M.} \bsnm{{Ogihara}}},
\bauthor{\binits{A.} \bsnm{{Morbidelli}}},
\bauthor{\binits{T.} \bsnm{{Guillot}}},
\batitle{{A reassessment of the in situ formation of close-in super-Earths}}.
\bjtitle{\aap}
\bvolume{578},
\bfpage{36}
(\byear{2015}).
doi:\doiurl{10.1051/0004-6361/201525884}
\end{barticle}
\endbibitem

\bibitem[\protect\citeauthoryear{{Ormel} and {Klahr}}{2010}]{Ormel10}
\begin{barticle}
\bauthor{\binits{C.W.} \bsnm{{Ormel}}},
\bauthor{\binits{H.H.} \bsnm{{Klahr}}},
\batitle{{The effect of gas drag on the growth of protoplanets. Analytical
  expressions for the accretion of small bodies in laminar disks}}.
\bjtitle{\aap}
\bvolume{520},
\bfpage{43}
(\byear{2010}).
doi:\doiurl{10.1051/0004-6361/201014903}
\end{barticle}
\endbibitem

\bibitem[\protect\citeauthoryear{{Owen} and {Murray-Clay}}{2018}]{Owen2018}
\begin{barticle}
\bauthor{\binits{J.E.} \bsnm{{Owen}}},
\bauthor{\binits{R.} \bsnm{{Murray-Clay}}},
\batitle{{Metallicity-dependent signatures in the Kepler planets}}.
\bjtitle{\mnras}
\bvolume{480}(\bissue{2}),
\bfpage{2206}--\blpage{2216}
(\byear{2018}).
doi:\doiurl{10.1093/mnras/sty1943}
\end{barticle}
\endbibitem

\bibitem[\protect\citeauthoryear{{Owen} and {Wu}}{2013}]{Owen2013}
\begin{barticle}
\bauthor{\binits{J.E.} \bsnm{{Owen}}},
\bauthor{\binits{Y.} \bsnm{{Wu}}},
\batitle{{Kepler Planets: A Tale of Evaporation}}.
\bjtitle{\apj}
\bvolume{775}(\bissue{2}),
\bfpage{105}
(\byear{2013}).
doi:\doiurl{10.1088/0004-637X/775/2/105}
\end{barticle}
\endbibitem

\bibitem[\protect\citeauthoryear{{Paardekooper} et~al.}{2013}]{Paardekooper13}
\begin{barticle}
\bauthor{\binits{S.-J.} \bsnm{{Paardekooper}}},
\bauthor{\binits{H.} \bsnm{{Rein}}},
\bauthor{\binits{W.} \bsnm{{Kley}}},
\batitle{{The formation of systems with closely spaced low-mass planets and the
  application to Kepler-36}}.
\bjtitle{\mnras}
\bvolume{434},
\bfpage{3018}--\blpage{3029}
(\byear{2013}).
doi:\doiurl{10.1093/mnras/stt1224}
\end{barticle}
\endbibitem

\bibitem[\protect\citeauthoryear{{Palle} et~al.}{2020}]{Palle2020}
\begin{barticle}
\bauthor{\binits{E.} \bsnm{{Palle}}},
\bauthor{\binits{L.} \bsnm{{Nortmann}}},
\bauthor{\binits{N.} \bsnm{{Casasayas-Barris}}},
\bauthor{\binits{M.} \bsnm{{Lamp{\'o}n}}},
\bauthor{\binits{M.} \bsnm{{L{\'o}pez-Puertas}}},
\bauthor{\binits{J.A.} \bsnm{{Caballero}}},
\bauthor{\binits{J.} \bsnm{{Sanz-Forcada}}},
\bauthor{\binits{L.M.} \bsnm{{Lara}}},
\bauthor{\binits{E.} \bsnm{{Nagel}}},
\bauthor{\binits{F.} \bsnm{{Yan}}},
\bauthor{\binits{F.J.} \bsnm{{Alonso-Floriano}}},
\bauthor{\binits{P.J.} \bsnm{{Amado}}},
\bauthor{\binits{G.} \bsnm{{Chen}}},
\bauthor{\binits{C.} \bsnm{{Cifuentes}}},
\bauthor{\binits{M.} \bsnm{{Cort{\'e}s-Contreras}}},
\bauthor{\binits{S.} \bsnm{{Czesla}}},
\bauthor{\binits{K.} \bsnm{{Molaverdikhani}}},
\bauthor{\binits{D.} \bsnm{{Montes}}},
\bauthor{\binits{V.M.} \bsnm{{Passegger}}},
\bauthor{\binits{A.} \bsnm{{Quirrenbach}}},
\bauthor{\binits{A.} \bsnm{{Reiners}}},
\bauthor{\binits{I.} \bsnm{{Ribas}}},
\bauthor{\binits{A.} \bsnm{{S{\'a}nchez-L{\'o}pez}}},
\bauthor{\binits{A.} \bsnm{{Schweitzer}}},
\bauthor{\binits{M.} \bsnm{{Stangret}}},
\bauthor{\binits{M.R.} \bsnm{{Zapatero Osorio}}},
\bauthor{\binits{M.} \bsnm{{Zechmeister}}},
\batitle{{A He I upper atmosphere around the warm Neptune GJ 3470 b}}.
\bjtitle{\aap}
\bvolume{638},
\bfpage{61}
(\byear{2020}).
doi:\doiurl{10.1051/0004-6361/202037719}
\end{barticle}
\endbibitem

\bibitem[\protect\citeauthoryear{{Papaloizou}}{2011}]{Papaloizou11}
\begin{barticle}
\bauthor{\binits{J.C.B.} \bsnm{{Papaloizou}}},
\batitle{{Tidal interactions in multi-planet systems}}.
\bjtitle{Celestial Mechanics and Dynamical Astronomy}
\bvolume{111},
\bfpage{83}--\blpage{103}
(\byear{2011}).
doi:\doiurl{10.1007/s10569-011-9344-4}
\end{barticle}
\endbibitem

\bibitem[\protect\citeauthoryear{Parmentier et~al.}{2013}]{Parmentier2013}
\begin{barticle}
\bauthor{\binits{V.} \bsnm{Parmentier}},
\bauthor{\binits{A.P.} \bsnm{Showman}},
\bauthor{\binits{Y.} \bsnm{Lian}},
\batitle{3d mixing in hot jupiters atmospheres-i. application to the day/night
  cold trap in hd 209458b}.
\bjtitle{Astronomy \& Astrophysics}
\bvolume{558},
\bfpage{91}
(\byear{2013})
\end{barticle}
\endbibitem

\bibitem[\protect\citeauthoryear{Parmentier et~al.}{2018}]{Parmentier2018}
\begin{barticle}
\bauthor{\binits{V.} \bsnm{Parmentier}},
\bauthor{\binits{M.R.} \bsnm{Line}},
\bauthor{\binits{J.L.} \bsnm{Bean}},
\bauthor{\binits{M.} \bsnm{Mansfield}},
\bauthor{\binits{L.} \bsnm{Kreidberg}},
\bauthor{\binits{R.} \bsnm{Lupu}},
\bauthor{\binits{C.} \bsnm{Visscher}},
\bauthor{\binits{J.-M.} \bsnm{D{\'e}sert}},
\bauthor{\binits{J.J.} \bsnm{Fortney}},
\bauthor{\binits{M.} \bsnm{Deleuil}}, \betal,
\batitle{From thermal dissociation to condensation in the atmospheres of ultra
  hot jupiters: Wasp-121b in context}.
\bjtitle{Astronomy \& Astrophysics}
\bvolume{617},
\bfpage{110}
(\byear{2018})
\end{barticle}
\endbibitem

\bibitem[\protect\citeauthoryear{{Penny} et~al.}{2019}]{Penny2019}
\begin{barticle}
\bauthor{\binits{M.T.} \bsnm{{Penny}}},
\bauthor{\binits{B.S.} \bsnm{{Gaudi}}},
\bauthor{\binits{E.} \bsnm{{Kerins}}},
\bauthor{\binits{N.J.} \bsnm{{Rattenbury}}},
\bauthor{\binits{S.} \bsnm{{Mao}}},
\bauthor{\binits{A.C.} \bsnm{{Robin}}},
\bauthor{\binits{S.} \bsnm{{Calchi Novati}}},
\batitle{{Predictions of the WFIRST Microlensing Survey. I. Bound Planet
  Detection Rates}}.
\bjtitle{\apjs}
\bvolume{241}(\bissue{1}),
\bfpage{3}
(\byear{2019}).
doi:\doiurl{10.3847/1538-4365/aafb69}
\end{barticle}
\endbibitem

\bibitem[\protect\citeauthoryear{{Petigura} et~al.}{2018}]{Petigura2018}
\begin{barticle}
\bauthor{\binits{E.A.} \bsnm{{Petigura}}},
\bauthor{\binits{G.W.} \bsnm{{Marcy}}},
\bauthor{\binits{J.N.} \bsnm{{Winn}}},
\bauthor{\binits{L.M.} \bsnm{{Weiss}}},
\bauthor{\binits{B.J.} \bsnm{{Fulton}}},
\bauthor{\binits{A.W.} \bsnm{{Howard}}},
\bauthor{\binits{E.} \bsnm{{Sinukoff}}},
\bauthor{\binits{H.} \bsnm{{Isaacson}}},
\bauthor{\binits{T.D.} \bsnm{{Morton}}},
\bauthor{\binits{J.A.} \bsnm{{Johnson}}},
\batitle{{The California-Kepler Survey. IV. Metal-rich Stars Host a Greater
  Diversity of Planets}}.
\bjtitle{\aj}
\bvolume{155}(\bissue{2}),
\bfpage{89}
(\byear{2018}).
doi:\doiurl{10.3847/1538-3881/aaa54c}
\end{barticle}
\endbibitem

\bibitem[\protect\citeauthoryear{{Pierens} et~al.}{2011}]{Pierens11}
\begin{barticle}
\bauthor{\binits{A.} \bsnm{{Pierens}}},
\bauthor{\binits{C.} \bsnm{{Baruteau}}},
\bauthor{\binits{F.} \bsnm{{Hersant}}},
\batitle{{On the dynamics of resonant super-Earths in disks with turbulence
  driven by stochastic forcing}}.
\bjtitle{\aap}
\bvolume{531},
\bfpage{5}
(\byear{2011}).
doi:\doiurl{10.1051/0004-6361/201116611}
\end{barticle}
\endbibitem

\bibitem[\protect\citeauthoryear{{Poleski} et~al.}{2014}]{Poleski2014}
\begin{barticle}
\bauthor{\binits{R.} \bsnm{{Poleski}}},
\bauthor{\binits{J.} \bsnm{{Skowron}}},
\bauthor{\binits{A.} \bsnm{{Udalski}}},
\bauthor{\binits{C.} \bsnm{{Han}}},
\bauthor{\binits{S.} \bsnm{{Koz{\l}owski}}},
\bauthor{\binits{{\L}.} \bsnm{{Wyrzykowski}}},
\bauthor{\binits{S.} \bsnm{{Dong}}},
\bauthor{\binits{M.K.} \bsnm{{Szyma{\'n}ski}}},
\bauthor{\binits{M.} \bsnm{{Kubiak}}},
\bauthor{\binits{G.} \bsnm{{Pietrzy{\'n}ski}}},
\bauthor{\binits{I.} \bsnm{{Soszy{\'n}ski}}},
\bauthor{\binits{K.} \bsnm{{Ulaczyk}}},
\bauthor{\binits{P.} \bsnm{{Pietrukowicz}}},
\bauthor{\binits{A.} \bsnm{{Gould}}},
\batitle{{Triple Microlens OGLE-2008-BLG-092L: Binary Stellar System with a
  Circumprimary Uranus-type Planet}}.
\bjtitle{\apj}
\bvolume{795}(\bissue{1}),
\bfpage{42}
(\byear{2014}).
doi:\doiurl{10.1088/0004-637X/795/1/42}
\end{barticle}
\endbibitem

\bibitem[\protect\citeauthoryear{{Poleski} et~al.}{2018}]{Poleski2018}
\begin{barticle}
\bauthor{\binits{R.} \bsnm{{Poleski}}},
\bauthor{\binits{B.S.} \bsnm{{Gaudi}}},
\bauthor{\binits{A.} \bsnm{{Udalski}}},
\bauthor{\binits{M.K.} \bsnm{{Szyma{\'n}ski}}},
\bauthor{\binits{I.} \bsnm{{Soszy{\'n}ski}}},
\bauthor{\binits{P.} \bsnm{{Pietrukowicz}}},
\bauthor{\binits{S.} \bsnm{{Koz{\l}owski}}},
\bauthor{\binits{J.} \bsnm{{Skowron}}},
\bauthor{\binits{{\L}.} \bsnm{{Wyrzykowski}}},
\bauthor{\binits{K.} \bsnm{{Ulaczyk}}},
\batitle{{An Ice Giant Exoplanet Interpretation of the Anomaly in Microlensing
  Event OGLE-2011-BLG-0173}}.
\bjtitle{\aj}
\bvolume{156}(\bissue{3}),
\bfpage{104}
(\byear{2018}).
doi:\doiurl{10.3847/1538-3881/aad45e}
\end{barticle}
\endbibitem

\bibitem[\protect\citeauthoryear{{Raymond} et~al.}{2014}]{Raymond2014}
\begin{bchapter}
\bauthor{\binits{S.N.} \bsnm{{Raymond}}},
\bauthor{\binits{E.} \bsnm{{Kokubo}}},
\bauthor{\binits{A.} \bsnm{{Morbidelli}}},
\bauthor{\binits{R.} \bsnm{{Morishima}}},
\bauthor{\binits{K.J.} \bsnm{{Walsh}}},
\bctitle{{Terrestrial Planet Formation at Home and Abroad}},
in \bbtitle{Protostars and Planets VI},
ed. by \beditor{\binits{H.} \bsnm{{Beuther}}},
\beditor{\binits{R.S.} \bsnm{{Klessen}}},
\beditor{\binits{C.P.} \bsnm{{Dullemond}}},
\beditor{\binits{T.} \bsnm{{Henning}}},
\byear{2014},
p. \bfpage{595}
\end{bchapter}
\endbibitem

\bibitem[\protect\citeauthoryear{{Rogers}}{2015}]{Rogers2015}
\begin{barticle}
\bauthor{\binits{L.A.} \bsnm{{Rogers}}},
\batitle{{Most 1.6 Earth-radius Planets are Not Rocky}}.
\bjtitle{\apj}
\bvolume{801}(\bissue{1}),
\bfpage{41}
(\byear{2015}).
doi:\doiurl{10.1088/0004-637X/801/1/41}
\end{barticle}
\endbibitem

\bibitem[\protect\citeauthoryear{{Rowan} et~al.}{2016}]{Rowan2016}
\begin{barticle}
\bauthor{\binits{D.} \bsnm{{Rowan}}},
\bauthor{\binits{S.} \bsnm{{Meschiari}}},
\bauthor{\binits{G.} \bsnm{{Laughlin}}},
\bauthor{\binits{S.S.} \bsnm{{Vogt}}},
\bauthor{\binits{R.P.} \bsnm{{Butler}}},
\bauthor{\binits{J.} \bsnm{{Burt}}},
\bauthor{\binits{S.} \bsnm{{Wang}}},
\bauthor{\binits{B.} \bsnm{{Holden}}},
\bauthor{\binits{R.} \bsnm{{Hanson}}},
\bauthor{\binits{P.} \bsnm{{Arriagada}}},
\bauthor{\binits{S.} \bsnm{{Keiser}}},
\bauthor{\binits{J.} \bsnm{{Teske}}},
\bauthor{\binits{M.} \bsnm{{Diaz}}},
\batitle{{The Lick-Carnegie Exoplanet Survey: HD 32963{\textemdash}A New
  Jupiter Analog Orbiting a Sun-like Star}}.
\bjtitle{\apj}
\bvolume{817}(\bissue{2}),
\bfpage{104}
(\byear{2016}).
doi:\doiurl{10.3847/0004-637X/817/2/104}
\end{barticle}
\endbibitem

\bibitem[\protect\citeauthoryear{{Santerne} et~al.}{2019}]{Santerne2019}
\begin{botherref}
\oauthor{\binits{A.} \bsnm{{Santerne}}},
\oauthor{\binits{L.} \bsnm{{Malavolta}}},
\oauthor{\binits{M.R.} \bsnm{{Kosiarek}}},
\oauthor{\binits{F.} \bsnm{{Dai}}},
\oauthor{\binits{C.D.} \bsnm{{Dressing}}},
\oauthor{\binits{X.} \bsnm{{Dumusque}}},
\oauthor{\binits{N.C.} \bsnm{{Hara}}},
\oauthor{\binits{T.A.} \bsnm{{Lopez}}},
\oauthor{\binits{A.} \bsnm{{Mortier}}},
\oauthor{\binits{A.} \bsnm{{Vanderburg}}},
\oauthor{\binits{V.} \bsnm{{Adibekyan}}},
\oauthor{\binits{D.J.} \bsnm{{Armstrong}}},
\oauthor{\binits{D.} \bsnm{{Barrado}}},
\oauthor{\binits{S.C.C.} \bsnm{{Barros}}},
\oauthor{\binits{D.} \bsnm{{Bayliss}}},
\oauthor{\binits{D.} \bsnm{{Berardo}}},
\oauthor{\binits{I.} \bsnm{{Boisse}}},
\oauthor{\binits{A.S.} \bsnm{{Bonomo}}},
\oauthor{\binits{F.} \bsnm{{Bouchy}}},
\oauthor{\binits{D.J.A.} \bsnm{{Brown}}},
\oauthor{\binits{L.A.} \bsnm{{Buchhave}}},
\oauthor{\binits{R.P.} \bsnm{{Butler}}},
\oauthor{\binits{A.} \bsnm{{Collier Cameron}}},
\oauthor{\binits{R.} \bsnm{{Cosentino}}},
\oauthor{\binits{J.D.} \bsnm{{Crane}}},
\oauthor{\binits{I.J.M.} \bsnm{{Crossfield}}},
\oauthor{\binits{M.} \bsnm{{Damasso}}},
\oauthor{\binits{M.R.} \bsnm{{Deleuil}}},
\oauthor{\binits{E.} \bsnm{{Delgado Mena}}},
\oauthor{\binits{O.} \bsnm{{Demangeon}}},
\oauthor{\binits{R.F.} \bsnm{{D{\'\i}az}}},
\oauthor{\binits{J.-F.} \bsnm{{Donati}}},
\oauthor{\binits{P.} \bsnm{{Figueira}}},
\oauthor{\binits{B.J.} \bsnm{{Fulton}}},
\oauthor{\binits{A.} \bsnm{{Ghedina}}},
\oauthor{\binits{A.} \bsnm{{Harutyunyan}}},
\oauthor{\binits{G.} \bsnm{{H{\'e}brard}}},
\oauthor{\binits{L.A.} \bsnm{{Hirsch}}},
\oauthor{\binits{S.} \bsnm{{Hojjatpanah}}},
\oauthor{\binits{A.W.} \bsnm{{Howard}}},
\oauthor{\binits{H.} \bsnm{{Isaacson}}},
\oauthor{\binits{D.W.} \bsnm{{Latham}}},
\oauthor{\binits{J.} \bsnm{{Lillo-Box}}},
\oauthor{\binits{M.} \bsnm{{L{\'o}pez-Morales}}},
\oauthor{\binits{C.} \bsnm{{Lovis}}},
\oauthor{\binits{A.F.} \bsnm{{Martinez Fiorenzano}}},
\oauthor{\binits{E.} \bsnm{{Molinari}}},
\oauthor{\binits{O.} \bsnm{{Mousis}}},
\oauthor{\binits{C.} \bsnm{{Moutou}}},
\oauthor{\binits{C.} \bsnm{{Nava}}},
\oauthor{\binits{L.D.} \bsnm{{Nielsen}}},
\oauthor{\binits{H.P.} \bsnm{{Osborn}}},
\oauthor{\binits{E.A.} \bsnm{{Petigura}}},
\oauthor{\binits{D.F.} \bsnm{{Phillips}}},
\oauthor{\binits{D.L.} \bsnm{{Pollacco}}},
\oauthor{\binits{E.} \bsnm{{Poretti}}},
\oauthor{\binits{K.} \bsnm{{Rice}}},
\oauthor{\binits{N.C.} \bsnm{{Santos}}},
\oauthor{\binits{D.} \bsnm{{S{\'e}gransan}}},
\oauthor{\binits{S.A.} \bsnm{{Shectman}}},
\oauthor{\binits{E.} \bsnm{{Sinukoff}}},
\oauthor{\binits{S.G.} \bsnm{{Sousa}}},
\oauthor{\binits{A.} \bsnm{{Sozzetti}}},
\oauthor{\binits{J.K.} \bsnm{{Teske}}},
\oauthor{\binits{S.} \bsnm{{Udry}}},
\oauthor{\binits{A.} \bsnm{{Vigan}}},
\oauthor{\binits{S.X.} \bsnm{{Wang}}},
\oauthor{\binits{C.A.} \bsnm{{Watson}}},
\oauthor{\binits{L.M.} \bsnm{{Weiss}}},
\oauthor{\binits{P.J.} \bsnm{{Wheatley}}},
\oauthor{\binits{J.N.} \bsnm{{Winn}}},
{An extremely low-density and temperate giant exoplanet}.
arXiv e-prints,
1911--07355
(2019)
\end{botherref}
\endbibitem

\bibitem[\protect\citeauthoryear{{Spiegel} et~al.}{2010}]{Spiegel2010}
\begin{barticle}
\bauthor{\binits{D.S.} \bsnm{{Spiegel}}},
\bauthor{\binits{A.} \bsnm{{Burrows}}},
\bauthor{\binits{L.} \bsnm{{Ibgui}}},
\bauthor{\binits{I.} \bsnm{{Hubeny}}},
\bauthor{\binits{J.A.} \bsnm{{Milsom}}},
\batitle{{Models of Neptune-Mass Exoplanets: Emergent Fluxes and Albedos}}.
\bjtitle{\apj}
\bvolume{709}(\bissue{1}),
\bfpage{149}--\blpage{158}
(\byear{2010}).
doi:\doiurl{10.1088/0004-637X/709/1/149}
\end{barticle}
\endbibitem

\bibitem[\protect\citeauthoryear{{Steffen} and {Hwang}}{2015}]{Steffen2015}
\begin{barticle}
\bauthor{\binits{J.H.} \bsnm{{Steffen}}},
\bauthor{\binits{J.A.} \bsnm{{Hwang}}},
\batitle{{The period ratio distribution of Kepler's candidate multiplanet
  systems}}.
\bjtitle{\mnras}
\bvolume{448}(\bissue{2}),
\bfpage{1956}--\blpage{1972}
(\byear{2015}).
doi:\doiurl{10.1093/mnras/stv104}
\end{barticle}
\endbibitem

\bibitem[\protect\citeauthoryear{{Tsiaras} et~al.}{2019}]{Tsiaras2019}
\begin{barticle}
\bauthor{\binits{A.} \bsnm{{Tsiaras}}},
\bauthor{\binits{I.P.} \bsnm{{Waldmann}}},
\bauthor{\binits{G.} \bsnm{{Tinetti}}},
\bauthor{\binits{J.} \bsnm{{Tennyson}}},
\bauthor{\binits{S.N.} \bsnm{{Yurchenko}}},
\batitle{{Water vapour in the atmosphere of the habitable-zone eight-Earth-mass
  planet K2-18 b}}.
\bjtitle{Nature Astronomy}
\bvolume{3},
\bfpage{1086}--\blpage{1091}
(\byear{2019}).
doi:\doiurl{10.1038/s41550-019-0878-9}
\end{barticle}
\endbibitem

\bibitem[\protect\citeauthoryear{{Van Eylen} et~al.}{2018}]{VanEylen2018}
\begin{barticle}
\bauthor{\binits{V.} \bsnm{{Van Eylen}}},
\bauthor{\binits{C.} \bsnm{{Agentoft}}},
\bauthor{\binits{M.S.} \bsnm{{Lundkvist}}},
\bauthor{\binits{H.} \bsnm{{Kjeldsen}}},
\bauthor{\binits{J.E.} \bsnm{{Owen}}},
\bauthor{\binits{B.J.} \bsnm{{Fulton}}},
\bauthor{\binits{E.} \bsnm{{Petigura}}},
\bauthor{\binits{I.} \bsnm{{Snellen}}},
\batitle{{An asteroseismic view of the radius valley: stripped cores, not born
  rocky}}.
\bjtitle{\mnras}
\bvolume{479}(\bissue{4}),
\bfpage{4786}--\blpage{4795}
(\byear{2018}).
doi:\doiurl{10.1093/mnras/sty1783}
\end{barticle}
\endbibitem

\bibitem[\protect\citeauthoryear{{Venturini} and {Helled}}{2017}]{Venturini17}
\begin{barticle}
\bauthor{\binits{J.} \bsnm{{Venturini}}},
\bauthor{\binits{R.} \bsnm{{Helled}}},
\batitle{{The Formation of Mini-Neptunes}}.
\bjtitle{\apj}
\bvolume{848}(\bissue{2}),
\bfpage{95}
(\byear{2017}).
doi:\doiurl{10.3847/1538-4357/aa8cd0}
\end{barticle}
\endbibitem

\bibitem[\protect\citeauthoryear{{Wakeford} et~al.}{2018}]{Wakeford2018}
\begin{barticle}
\bauthor{\binits{H.R.} \bsnm{{Wakeford}}},
\bauthor{\binits{D.K.} \bsnm{{Sing}}},
\bauthor{\binits{D.} \bsnm{{Deming}}},
\bauthor{\binits{N.K.} \bsnm{{Lewis}}},
\bauthor{\binits{J.} \bsnm{{Goyal}}},
\bauthor{\binits{T.J.} \bsnm{{Wilson}}},
\bauthor{\binits{J.} \bsnm{{Barstow}}},
\bauthor{\binits{T.} \bsnm{{Kataria}}},
\bauthor{\binits{B.} \bsnm{{Drummond}}},
\bauthor{\binits{T.M.} \bsnm{{Evans}}},
\bauthor{\binits{A.L.} \bsnm{{Carter}}},
\bauthor{\binits{N.} \bsnm{{Nikolov}}},
\bauthor{\binits{H.A.} \bsnm{{Knutson}}},
\bauthor{\binits{G.E.} \bsnm{{Ballester}}},
\bauthor{\binits{A.M.} \bsnm{{Mand ell}}},
\batitle{{The Complete Transmission Spectrum of WASP-39b with a Precise Water
  Constraint}}.
\bjtitle{\aj}
\bvolume{155}(\bissue{1}),
\bfpage{29}
(\byear{2018}).
doi:\doiurl{10.3847/1538-3881/aa9e4e}
\end{barticle}
\endbibitem

\bibitem[\protect\citeauthoryear{{Wakeford} et~al.}{2017}]{Wakeford2017}
\begin{barticle}
\bauthor{\binits{H.R.} \bsnm{{Wakeford}}},
\bauthor{\binits{D.K.} \bsnm{{Sing}}},
\bauthor{\binits{T.} \bsnm{{Kataria}}},
\bauthor{\binits{D.} \bsnm{{Deming}}},
\bauthor{\binits{N.} \bsnm{{Nikolov}}},
\bauthor{\binits{E.D.} \bsnm{{Lopez}}},
\bauthor{\binits{P.} \bsnm{{Tremblin}}},
\bauthor{\binits{D.S.} \bsnm{{Amundsen}}},
\bauthor{\binits{N.K.} \bsnm{{Lewis}}},
\bauthor{\binits{A.M.} \bsnm{{Mandell}}},
\bauthor{\binits{J.J.} \bsnm{{Fortney}}},
\bauthor{\binits{H.} \bsnm{{Knutson}}},
\bauthor{\binits{B.} \bsnm{{Benneke}}},
\bauthor{\binits{T.M.} \bsnm{{Evans}}},
\batitle{{HAT-P-26b: A Neptune-mass exoplanet with a well-constrained heavy
  element abundance}}.
\bjtitle{Science}
\bvolume{356}(\bissue{6338}),
\bfpage{628}--\blpage{631}
(\byear{2017}).
doi:\doiurl{10.1126/science.aah4668}
\end{barticle}
\endbibitem

\bibitem[\protect\citeauthoryear{{Weiss} and {Marcy}}{2014}]{Weiss2014}
\begin{barticle}
\bauthor{\binits{L.M.} \bsnm{{Weiss}}},
\bauthor{\binits{G.W.} \bsnm{{Marcy}}},
\batitle{{The Mass-Radius Relation for 65 Exoplanets Smaller than 4 Earth
  Radii}}.
\bjtitle{\apjl}
\bvolume{783}(\bissue{1}),
\bfpage{6}
(\byear{2014}).
doi:\doiurl{10.1088/2041-8205/783/1/L6}
\end{barticle}
\endbibitem

\bibitem[\protect\citeauthoryear{{Weiss} et~al.}{2013}]{Weiss2013}
\begin{barticle}
\bauthor{\binits{L.M.} \bsnm{{Weiss}}},
\bauthor{\binits{G.W.} \bsnm{{Marcy}}},
\bauthor{\binits{J.F.} \bsnm{{Rowe}}},
\bauthor{\binits{A.W.} \bsnm{{Howard}}},
\bauthor{\binits{H.} \bsnm{{Isaacson}}},
\bauthor{\binits{J.J.} \bsnm{{Fortney}}},
\bauthor{\binits{N.} \bsnm{{Miller}}},
\bauthor{\binits{B.-O.} \bsnm{{Demory}}},
\bauthor{\binits{D.A.} \bsnm{{Fischer}}},
\bauthor{\binits{E.R.} \bsnm{{Adams}}},
\bauthor{\binits{A.K.} \bsnm{{Dupree}}},
\bauthor{\binits{S.B.} \bsnm{{Howell}}},
\bauthor{\binits{R.} \bsnm{{Kolbl}}},
\bauthor{\binits{J.A.} \bsnm{{Johnson}}},
\bauthor{\binits{E.P.} \bsnm{{Horch}}},
\bauthor{\binits{M.E.} \bsnm{{Everett}}},
\bauthor{\binits{D.C.} \bsnm{{Fabrycky}}},
\bauthor{\binits{S.} \bsnm{{Seager}}},
\batitle{{The Mass of KOI-94d and a Relation for Planet Radius, Mass, and
  Incident Flux}}.
\bjtitle{\apj}
\bvolume{768}(\bissue{1}),
\bfpage{14}
(\byear{2013}).
doi:\doiurl{10.1088/0004-637X/768/1/14}
\end{barticle}
\endbibitem

\bibitem[\protect\citeauthoryear{Welbanks et~al.}{2019}]{Welbanks2019}
\begin{barticle}
\bauthor{\binits{L.} \bsnm{Welbanks}},
\bauthor{\binits{N.} \bsnm{Madhusudhan}},
\bauthor{\binits{N.F.} \bsnm{Allard}},
\bauthor{\binits{I.} \bsnm{Hubeny}},
\bauthor{\binits{F.} \bsnm{Spiegelman}},
\bauthor{\binits{T.} \bsnm{Leininger}},
\batitle{Mass--metallicity trends in transiting exoplanets from atmospheric
  abundances of h2o, na, and k}.
\bjtitle{The Astrophysical Journal Letters}
\bvolume{887}(\bissue{1}),
\bfpage{20}
(\byear{2019})
\end{barticle}
\endbibitem

\bibitem[\protect\citeauthoryear{{Wittenmyer} et~al.}{2014}]{Wittenmyer2014}
\begin{barticle}
\bauthor{\binits{R.A.} \bsnm{{Wittenmyer}}},
\bauthor{\binits{J.} \bsnm{{Horner}}},
\bauthor{\binits{C.G.} \bsnm{{Tinney}}},
\bauthor{\binits{R.P.} \bsnm{{Butler}}},
\bauthor{\binits{H.R.A.} \bsnm{{Jones}}},
\bauthor{\binits{M.} \bsnm{{Tuomi}}},
\bauthor{\binits{G.S.} \bsnm{{Salter}}},
\bauthor{\binits{B.D.} \bsnm{{Carter}}},
\bauthor{\binits{F.E.} \bsnm{{Koch}}},
\bauthor{\binits{S.J.} \bsnm{{O'Toole}}},
\bauthor{\binits{J.} \bsnm{{Bailey}}},
\bauthor{\binits{D.} \bsnm{{Wright}}},
\batitle{{The Anglo-Australian Planet Search. XXIII. Two New Jupiter Analogs}}.
\bjtitle{\apj}
\bvolume{783}(\bissue{2}),
\bfpage{103}
(\byear{2014}).
doi:\doiurl{10.1088/0004-637X/783/2/103}
\end{barticle}
\endbibitem

\bibitem[\protect\citeauthoryear{{Wittenmyer} et~al.}{2016}]{Wittenmyer2016}
\begin{barticle}
\bauthor{\binits{R.A.} \bsnm{{Wittenmyer}}},
\bauthor{\binits{R.P.} \bsnm{{Butler}}},
\bauthor{\binits{C.G.} \bsnm{{Tinney}}},
\bauthor{\binits{J.} \bsnm{{Horner}}},
\bauthor{\binits{B.D.} \bsnm{{Carter}}},
\bauthor{\binits{D.J.} \bsnm{{Wright}}},
\bauthor{\binits{H.R.A.} \bsnm{{Jones}}},
\bauthor{\binits{J.} \bsnm{{Bailey}}},
\bauthor{\binits{S.J.} \bsnm{{O'Toole}}},
\batitle{{The Anglo-Australian Planet Search XXIV: The Frequency of Jupiter
  Analogs}}.
\bjtitle{\apj}
\bvolume{819}(\bissue{1}),
\bfpage{28}
(\byear{2016}).
doi:\doiurl{10.3847/0004-637X/819/1/28}
\end{barticle}
\endbibitem

\bibitem[\protect\citeauthoryear{{Wong} et~al.}{2020}]{Wong2020}
\begin{barticle}
\bauthor{\binits{I.} \bsnm{{Wong}}},
\bauthor{\binits{B.} \bsnm{{Benneke}}},
\bauthor{\binits{P.} \bsnm{{Gao}}},
\bauthor{\binits{H.A.} \bsnm{{Knutson}}},
\bauthor{\binits{Y.} \bsnm{{Chachan}}},
\bauthor{\binits{G.W.} \bsnm{{Henry}}},
\bauthor{\binits{D.} \bsnm{{Deming}}},
\bauthor{\binits{T.} \bsnm{{Kataria}}},
\bauthor{\binits{G.K.H.} \bsnm{{Lee}}},
\bauthor{\binits{N.} \bsnm{{Nikolov}}},
\bauthor{\binits{D.K.} \bsnm{{Sing}}},
\bauthor{\binits{G.E.} \bsnm{{Ballester}}},
\bauthor{\binits{N.J.} \bsnm{{Baskin}}},
\bauthor{\binits{H.R.} \bsnm{{Wakeford}}},
\bauthor{\binits{M.H.} \bsnm{{Williamson}}},
\batitle{{Optical to Near-infrared Transmission Spectrum of the Warm Sub-Saturn
  HAT-P-12b}}.
\bjtitle{\aj}
\bvolume{159}(\bissue{5}),
\bfpage{234}
(\byear{2020}).
doi:\doiurl{10.3847/1538-3881/ab880d}
\end{barticle}
\endbibitem

\bibitem[\protect\citeauthoryear{{Zeng} et~al.}{2019}]{Zeng2019}
\begin{barticle}
\bauthor{\binits{L.} \bsnm{{Zeng}}},
\bauthor{\binits{S.B.} \bsnm{{Jacobsen}}},
\bauthor{\binits{D.D.} \bsnm{{Sasselov}}},
\bauthor{\binits{M.I.} \bsnm{{Petaev}}},
\bauthor{\binits{A.} \bsnm{{Vanderburg}}},
\bauthor{\binits{M.} \bsnm{{Lopez-Morales}}},
\bauthor{\binits{J.} \bsnm{{Perez-Mercader}}},
\bauthor{\binits{T.R.} \bsnm{{Mattsson}}},
\bauthor{\binits{G.} \bsnm{{Li}}},
\bauthor{\binits{M.Z.} \bsnm{{Heising}}},
\bauthor{\binits{A.S.} \bsnm{{Bonomo}}},
\bauthor{\binits{M.} \bsnm{{Damasso}}},
\bauthor{\binits{T.A.} \bsnm{{Berger}}},
\bauthor{\binits{H.} \bsnm{{Cao}}},
\bauthor{\binits{A.} \bsnm{{Levi}}},
\bauthor{\binits{R.D.} \bsnm{{Wordsworth}}},
\batitle{{Growth model interpretation of planet size distribution}}.
\bjtitle{Proceedings of the National Academy of Science}
\bvolume{116}(\bissue{20}),
\bfpage{9723}--\blpage{9728}
(\byear{2019}).
doi:\doiurl{10.1073/pnas.1812905116}
\end{barticle}
\endbibitem

\bibitem[\protect\citeauthoryear{{Zhang} and {Showman}}{2018}]{Zhang2018}
\begin{barticle}
\bauthor{\binits{X.} \bsnm{{Zhang}}},
\bauthor{\binits{A.P.} \bsnm{{Showman}}},
\batitle{{Global-mean Vertical Tracer Mixing in Planetary Atmospheres. I.
  Theory and Fast-rotating Planets}}.
\bjtitle{\apj}
\bvolume{866}(\bissue{1}),
\bfpage{1}
(\byear{2018}).
doi:\doiurl{10.3847/1538-4357/aada85}
\end{barticle}
\endbibitem

\bibitem[\protect\citeauthoryear{{Zhu} et~al.}{2018}]{Zhu2018}
\begin{barticle}
\bauthor{\binits{W.} \bsnm{{Zhu}}},
\bauthor{\binits{C.} \bsnm{{Petrovich}}},
\bauthor{\binits{Y.} \bsnm{{Wu}}},
\bauthor{\binits{S.} \bsnm{{Dong}}},
\bauthor{\binits{J.} \bsnm{{Xie}}},
\batitle{{About 30\% of Sun-like Stars Have Kepler-like Planetary Systems: A
  Study of Their Intrinsic Architecture}}.
\bjtitle{\apj}
\bvolume{860}(\bissue{2}),
\bfpage{101}
(\byear{2018}).
doi:\doiurl{10.3847/1538-4357/aac6d5}
\end{barticle}
\endbibitem

\end{thebibliography}

%
%

\end{document}